\documentclass[11pt]{article}%
\usepackage{amsmath}
\usepackage{graphicx}%
\usepackage{amsfonts}%
\usepackage{amssymb}
\newcommand{\eqnb}{\begin{equation}}
\newcommand{\eqne}{\end{equation}}

\newtheorem{The}{Theorem}

\newtheorem{Cor}[The]{Corollary}
\newtheorem{Lem}{Lemma}
\newtheorem{Pro}{Proposition}
\newtheorem{Rem}{Remark}

\begin{document}

\title{\textbf{Doubly Exponential Solution for Randomized Load Balancing Models with
General Service Times}}
\author{Quan-Lin Li\\School of Economics and Management Sciences \\Yanshan University, Qinhuangdao 066004, P.R. China}
\date{September 5, 2010}
\maketitle

\begin{abstract}
The randomized load balancing model (also called \emph{supermarket model}) is
now being applied to the study of load balancing in data centers and
multi-core servers systems. It is very interesting to analyze the general
service times in the supermarket model, and specifically understand influence
of the heavy-tailed service times on the doubly exponential solution. Since
the supermarket model is a complex queueing system, the general service times
make its analysis more challenging than the exponential or PH service case. Up
to now, it still is an open problem whether or how the heavy-tailed service
times can disrupt the doubly exponential structure of the fixed point in the
supermarket model.

In this paper, we provide a novel and simple approach to study the supermarket
model with general service times. This approach is based on the supplementary
variable method used in analyzing stochastic models extensively. We organize
an infinite-size system of integral-differential equations by means of the
density dependent jump Markov process, and obtain a close-form solution:
doubly exponential structure, for the fixed point satisfying the system of
nonlinear equations, which is always a key in the study of supermarket models.
The fixed point is decomposited into two groups of information under a product
form: the arrival information and the service information. Based on this, we
indicate two important observations: the fixed point for the supermarket model
is different from the tail of stationary queue length distribution for the
ordinary M/G/1 queue, and the doubly exponential solution to the fixed point
can extensively exist even if the service time distribution is heavy-tailed.
Furthermore, we analyze the exponential convergence of the current location of
the supermarket model to its fixed point, and study the Lipschitz condition in
the Kurtz Theorem under general service times. Based on these analysis, one
can gain a new understanding how workload probing can help in load balancing
jobs with general service times such as heavy-tailed service.

\vskip                                                      0.5cm

\noindent\textbf{Keywords:} Randomized load balancing, supermarket model,
density dependent jump Markov process, fixed point, doubly exponential
solution, heavy-tailed distribution, exponential convergence, Lipschitz condition.

\end{abstract}

\section{Introduction}

Randomized load balancing, where a job is assigned to a server from a small
subset of randomly chosen servers, is very simple to implement, and can
surprisingly deliver better performance (for example reducing collisions,
waiting times, backlogs) in a number of applications, such as, data center,
hash tables, distributed memory machines, path selection in networks, and task
assignment at web servers. One useful model that has been extensively used to
study the randomized load balancing schemes is the supermarket model. In the
supermarket model, a key result by Vvedenskaya, Dobrushin and Karpelevich
\cite{Vve:1996} indicated that when each Poisson arriving job is assigned to
the shortest one of $d\geq2$ randomly chosen queues with exponential service
times, the equilibrium queue length can decay doubly exponentially in the
limit as the population number $n\rightarrow\infty$, and the stationary
fraction of queues with at least $k$ customers is $\rho^{\frac{d^{k}-1}{d-1}}%
$, which is a substantially exponential improvement over the case for $d=1$,
where the tail of stationary queue length distribution in the corresponding
M/M/1 queue is $\rho^{k}$.

The distributed load balancing strategies, in which individual job decisions
are based on information on a limited number of other processors, have been
studied analytically by Eager, Lazokwska and Zahorjan \cite{Eag:1986a,
Eag:1986b, Eag:1988} and through trace-driven simulations by Zhou
\cite{Zhou:1988}. Based on this, the supermarket model is developed by
queueing theory and Markov processes. Most of recent research applied the
density dependent jump Markov processes to deal with a simple supermarket
model with Poisson arrival processes and exponential service times, a key
result of which illustrates that there exists a unique fixed point which is
decreasing doubly exponentially. That approach used in the literature relies
on determining the behavior of the supermarket model as its size grows to
infinity, and its behavior is naturally described as a system of differential
equations, which leads to a closed form solution: doubly exponential
structure, of the fixed point. Readers may refer to, such as, analyzing a
basic and simple supermarket model by Azar, Broder, Karlin and Upfal
\cite{Azar:1999}, Vvedenskaya, Dobrushin and Karpelevich \cite{Vve:1996},
Mitzenmacher \cite{Mit:1996a, Mit:1996b}.

Certain generalization of the supermarket model have been explored, for
example, simple variations by Mitzenmacher and V\"{o}cking \cite{Mit:1998},
Mitzenmacher \cite{Mit:1998a, Mit:1999a, Mit:2001}, V\"{o}cking
\cite{Voc:1999}, Mitzenmacher, Richa, and Sitaraman \cite{Mit:2001a} and
Vvedenskaya and Suhov \cite{Vve:1997}; and analyzing load information by
Mirchandaney, Towsley, and Stankovic \cite{Mir:1989}, Dahlin \cite{Dah:1999},
Mitzenmacher \cite{Mit:2000, Mit:2001a}. Furthermore, Martin and Suhov
\cite{Mar:1999}, Martin \cite{Mar:2001}, Suhov and Vvedenskaya \cite{Suh:2002}
studied the supermarket mall model by means of the fast Jackson network, where
each node in a Jackson network is replaced by $N$ parallel servers, and a job
joins the shortest of $d$ randomly chosen queues at the node to which it is
directed. Luczak and McDiamid \cite{Luc:2005, Luc:2006} studied the maximum
queue length of the original supermarket model with exponential service times
when the service speed scales linearly with the number of jobs in the queue.
Li, Lui and Wang \cite{Li:2010a,Li:2010b} discussed the supermarket model with
PH service times and the supermarket model with Markovian arrival processes,
respectively. Readers may refer to an excellent overview by Mitzenmacher,
Richa, and Sitaraman \cite{Mit:2001a}.

This paper is interested in analyzing the supermarket model with general
service times, which is an open problem for determining whether or how the
heavy-tailed service times can disrupt the doubly exponential structure of the
fixed point. On the other hand, note that the supermarket model is a complex
queueing system and has much different characteristics from the ordinary
queueing systems, thus the general service times make its analysis more
challenging than the exponential or PH service case. Up to now, there has not
been an effective method to be able to deal with the supermarket model with
general service times.

The main contributions of the paper are threefold. The first one is to provide
a novel and simple approach to study the supermarket model with general
service times. This approach is based on the supplementary variable method but
is described as a new integral-differential structure for expressing and
computing the fraction of queues efficiently. Using the new approach, we setup
an infinite-size system of integral-differential equations, which makes
applications of the density dependent jump Markov processes to be able to deal
with the general distributions, such as general service times, involved in the
supermarket model. The second one is to obtain a close-form solution: doubly
exponential structure, for the fixed point satisfying the system of nonlinear
equations, which is always a key in the study of supermarket models.
Furthermore, this paper analyzes the exponential convergence of the current
location of the supermarket model to its fixed point, and studies the
Lipschitz condition in the Kurtz Theorem under general service times. Also,
this paper provides numerical examples to illustrate the effectiveness of our
approach in analyzing the randomized load balancing schemes with the
non-exponential service requirements. The third one is to obtain that the
fixed point is decomposited into two groups of information under a product
form: the arrival information and the service information. Based on this, we
indicate three important observations:

\begin{description}
\item[(a)] \noindent The fixed point for the supermarket model is different
from the tail of stationary queue length distribution for the ordinary M/G/1
queue, because the fixed point is light-tailed but the stationary queue length
is heavy-tail if the service times are heavy-tailed. Note that such a
difference is illustrated in this paper for the first time, while it can not
be observed in the literature for the supermarket model with Poisson arrivals
and exponential service times, e.g., see Mitzenmacher, Richa, and Sitaraman
\cite{Mit:2001a}.

\item[(b)] \noindent The doubly exponential solution to the fixed point can
extensively exist even if the service time distribution is heavy-tailed. This
is an answer of the above open problem to illustrate the role played by the
heavy-tailed service time distribution in the doubly exponential solution to
the fixed point.

\item[(c)] \noindent The doubly exponential solution to the fixed point is not
unique for a more general supermarket model. Note that we give three different
doubly exponential solutions in the supermarket model with Poisson arrivals
and PH service times, thus it is very interesting to provide all the doubly
exponential solutions for a more general supermarket model.
\end{description}

\noindent Based on this, one can gain the new and important understanding how
the workload probing can help in load balancing jobs with general service
times such as heavy-tailed service.

The remainder of this paper is organized as follows. In Section 2, we first
describe a supermarket model with general service times, which is always
useful in the study of randomized load balancing schemes. Then the supermarket
model is expressed as a systems of integral-differential equations in terms of
the density dependent jump Markov processes. In Section 3, we first introduce
a fixed point of the system of integral-differential equations, and set up a
system of nonlinear equations satisfied by the fixed point. Then we provide a
close-form solution: doubly exponential structure, to the system of nonlinear
equations. In Section 4, we provide a necessary discussion on the key
parameter $\theta$ used in the doubly exponential structure, and indicate that
the doubly exponential solution to the fixed point extensively exists even if
the service time distribution is heavy-tailed. In Section 5, we give three
methods to analyze the supermarket model with Poisson arrivals and PH service
times, and provide three different ways to determine the key parameter
$\theta$ and compute the doubly exponential solution to the fixed point. We
show that the doubly exponential solution to the fixed point is not unique for
a more general supermarket. In Section 6, we study the exponential convergence
of the current location of the supermarket model to its fixed point. Not only
does the exponential convergence indicates the existence of the fixed point,
but it also explains such a convergent process is very fast. In Section 7, we
apply the Kurtz Theorem to study the supermarket model with the general
service times, and analyze the Lipschitz condition with respect to general
service times. Some concluding remarks are given in Section 8.

\section{Supermarket Model}

In this section, we first describe a supermarket model with general service
times. Then we provide a novel and simple approach to setup an infinite-size
system of integral-differential equations based on the density dependent jump
Markov processes. Note that this approach is based on the supplementary
variable method but is described as a new integral-differential structure so
that the corresponding boundary conditions are written in a different version.

Let us formally describe the supermarket model, which is abstracted as a
multi-server multi-queue stochastic system. Customers arrive at a queueing
system of $n>1$ servers as a Poisson process with an average arrival rate
$n\lambda$ for $\lambda>0$. The service time $\chi_{k}$ of the $k$th customer
is general with the distribution function%
\[
G\left(  x\right)  =P\left\{  \chi_{k}\leq x\right\}  =1-\exp\left\{
-\int_{0}^{x}\mu\left(  y\right)  dy\right\}  ,
\]
where all the random variables $\chi_{k}$ for $k\geq1$ are i.i.d. with the
mean $E\left[  \chi_{k}\right]  =1/\mu$. Each arriving customer chooses
$d\geq1$ servers independently and uniformly at random from these $n$ servers,
and waits for service at the server which currently contains the fewest number
of customers. If there is a tie, servers with the fewest number of customers
will be chosen randomly. All customers in any service center will be served in
the First-Come-First-Served (FCFS) manner. Figure 1 simply shows such a
supermarket model.

\begin{figure}[ptbh]
\centering                     \includegraphics[width=10cm]{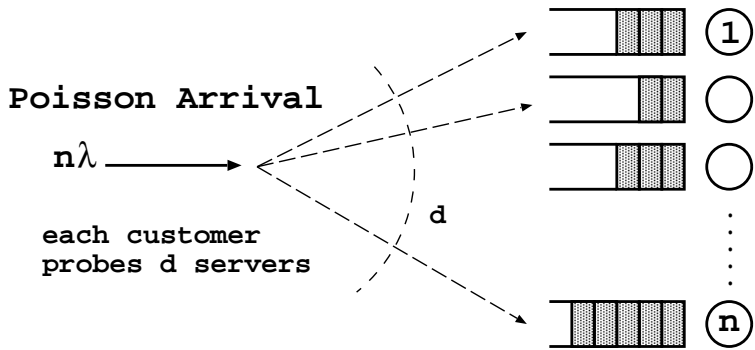}
\caption{working structure of the supermarket model}%
\label{figure: model}%
\end{figure}

In the study of supermarket models, it is necessary for us to study general
service time distributions, for example, heavy-tailed distributions. Not only
because the general distribution makes analysis of the supermarket models more
difficult and challenging than those in the literature for the exponential or
PH service case, but it also allows us to model more realistic systems and
understand their performance implication under the randomized load balancing
strategy. As indicated in \cite{harchol97}, the process times of many parallel
jobs, in particular, jobs to data centers, tend to be non-exponential. Unless
we state otherwise, we assume that all the random variables defined above are
independent, and that the system is operating under the condition:
$\rho=\lambda/\mu<1$.

\begin{Lem}
The supermarket model with general service times is stable if $\rho
=\lambda/\mu<1.$
\end{Lem}

\textbf{Proof:} Let $Q_{O}\left(  t\right)  $ and $Q_{S}\left(  t\right)  $ be
the queue lengths of the ordinary M/G/1 queue and of an arbitrary server in
the supermarket model at time $t$, respectively. Note that in the supermarket
model, each customer chooses $d$ servers independently and uniformily at
random, and the queue length of the entering server is currently shorten, it
is easy to see that for each $t\geq0$,%
\begin{equation}
0\leq Q_{S}\left(  t\right)  \underset{\text{st}}{\leq}Q_{O}\left(  t\right)
\text{.} \label{Equ0}%
\end{equation}
Since the ordinary M/G/1 queue is stable if $\rho=\lambda/\mu<1$, it follows
from (\ref{Equ0}) that the supermarket model with general service times is
stable if $\rho=\lambda/\mu<1$. This completes the proof. \hspace*{\fill} \rule{1.8mm}{2.5mm}

For $k\geq1$, we define $n_{k}\left(  t,x\right)  $d$x$ as the number of
queues with at least $k$ customers and the residual service time of each
server be in the interval $[x,x+$d$x)$ at time $t\geq0$. Clearly, $0\leq
n_{k}\left(  t,x\right)  \leq n$ for $x\geq0$ and $1\leq k\leq n$. Let%
\[
s_{k,n}\left(  t,x\right)  =\frac{n_{k}\left(  t,x\right)  }{n},
\]
which is the density function of the fraction of queues with at least $k$
customers and the residual service time of each server be $x$. We write%
\[
S_{k}\left(  t,x\right)  =\lim_{n\rightarrow\infty}s_{k,n}\left(  t,x\right)
,\text{ \ for }k\geq1.
\]
We define $n_{0,n}\left(  t\right)  $ as the number of queues with at least
$0$ customers at time $t\geq0$. Clearly, $n_{0,n}\left(  t\right)  =n$. Let%
\[
s_{0,n}\left(  t\right)  =\frac{n_{0,n}\left(  t\right)  }{n}.
\]
Then $s_{0,n}\left(  t\right)  =1$ for all $t\geq0$ and%
\[
S_{0}\left(  t\right)  =\lim_{n\rightarrow\infty}s_{0,n}\left(  t\right)  =1.
\]
Let $V\left(  t\right)  $ be the fraction of queues with zero customer at time
$t$. Then%
\[
S_{0}\left(  t\right)  =V\left(  t\right)  +\int_{0}^{+\infty}S_{1}\left(
t,x\right)  \text{d}x.
\]
Thus we have%
\[
S\left(  t,x\right)  =\left(  S_{0}\left(  t\right)  ,S_{1}\left(  t,x\right)
,S_{2}\left(  t,x\right)  ,\ldots\right)  .
\]

The following proposition shows that the sequence $\left\{  S_{k}\left(
t,x\right)  \right\}  $ is monotone increasing for $k\geq1$, while its proof
is easily by means of the definition of $S_{k}\left(  t,x\right)  $ for
$k\geq1$.

\begin{Pro}
For $1\leq k<l$%
\[
S_{l}\left(  t,x\right)  <S_{k}\left(  t,x\right)  <S_{0}\left(  t\right)  =1
\]
and%
\[
\int_{0}^{+\infty}S_{l}\left(  t,x\right)  \text{d}x<\int_{0}^{+\infty}%
S_{k}\left(  t,x\right)  \text{d}x<S_{0}\left(  t\right)  =1.
\]
\end{Pro}

Now, we setup a system of integral-differential equations by means of the
density dependent jump Markov process. To that end, we provide an example with
$k\geq2$ to indicate how to derive the system of integral-differential equations.

Consider the supermarket model with $n$ queues, and determine the expected
change in the number of servers with at least $k$ customers and the residual
service time of each server be $x$ over a small time period of length d$t$.
The probability that a customer arriving during this time period is $n\lambda
$d$t$, and the probability that an arriving customer joins a queue of size
$k-1$ is given by $\int_{0}^{+\infty}s_{k-1,n}^{d}\left(  t,x\right)
$d$x-\int_{0}^{+\infty}s_{k,n}^{d}\left(  t,x\right)  $d$x$. Thus, the
probability that during this time period, any arriving customer joins a queue
of size $k-1$ is given by%
\[
n\lambda\text{d}t\cdot\left[  \int_{0}^{+\infty}s_{k-1,n}^{d}\left(
t,x\right)  \text{d}x-\int_{0}^{+\infty}s_{k,n}^{d}\left(  t,x\right)
\text{d}x\right]  .
\]
Similarly, the probability that a customer leaves a server of size $k$ is
given by%
\[
ndt\cdot\left[  \int_{0}^{+\infty}\mu\left(  x\right)  s_{k,n}\left(
t,x\right)  \text{d}x-\int_{0}^{+\infty}\mu\left(  x\right)  s_{k+1,n}\left(
t,x\right)  \text{d}x\right]  .
\]
Therefore we can obtain%
\begin{align*}
\frac{\text{d}\int_{0}^{+\infty}n_{k}\left(  t,x\right)  \text{d}x}{\text{d}%
t}=  &  n\lambda\left[  \int_{0}^{+\infty}s_{k-1,n}^{d}\left(  t,x\right)
\text{d}x-\int_{0}^{+\infty}s_{k,n}^{d}\left(  t,x\right)  \text{d}x\right] \\
&  +n\left[  \int_{0}^{+\infty}\mu\left(  x\right)  s_{k,n}\left(  t,x\right)
\text{d}x-\int_{0}^{+\infty}\mu\left(  x\right)  s_{k+1,n}\left(  t,x\right)
\text{d}x\right]  ,
\end{align*}
which leads to%
\begin{align}
\frac{\text{d}\int_{0}^{+\infty}s_{k,n}\left(  t,x\right)  \text{d}x}%
{\text{d}t}=  &  \lambda\left[  \int_{0}^{+\infty}s_{k-1,n}^{d}\left(
t,x\right)  \text{d}x-\int_{0}^{+\infty}s_{k,n}^{d}\left(  t,x\right)
\text{d}x\right] \nonumber\\
&  +\left[  \int_{0}^{+\infty}\mu\left(  x\right)  s_{k,n}\left(  t,x\right)
\text{d}x-\int_{0}^{+\infty}\mu\left(  x\right)  s_{k+1,n}\left(  t,x\right)
\text{d}x\right]  . \label{Equ1}%
\end{align}
Taking $n\rightarrow\infty$ in the both sides of (\ref{Equ1}), we have%
\begin{align}
\frac{\text{d}\int_{0}^{+\infty}S_{k}\left(  t\right)  \text{d}x}{\text{d}t}=
&  \lambda\left[  \int_{0}^{+\infty}S_{k-1}^{d}\left(  t,x\right)
\text{d}x-\int_{0}^{+\infty}S_{k}^{d}\left(  t,x\right)  \text{d}x\right]
\nonumber\\
&  +\left[  \int_{0}^{+\infty}\mu\left(  x\right)  S_{k}\left(  t,x\right)
\text{d}x-\int_{0}^{+\infty}\mu\left(  x\right)  S_{k+1}\left(  t,x\right)
\text{d}x\right]  . \label{Equ2}%
\end{align}

Using a similar analysis to that for deriving Equation (\ref{Equ2}), we can
easily obtain a system of integral-differential equations for the fraction
density vector $S\left(  t,x\right)  $ as follows:%
\begin{equation}
S_{0}\left(  t\right)  =1\text{ for all }t\geq0, \label{Equ3}%
\end{equation}%
\begin{equation}
\frac{\mathtt{d}}{\text{d}t}S_{0}\left(  t\right)  =-\lambda S_{0}^{d}\left(
t\right)  +\int_{0}^{+\infty}\mu\left(  x\right)  S_{1}\left(  t,x\right)
\text{d}x, \label{Equ4}%
\end{equation}%
\begin{align}
\frac{\mathtt{d}\int_{0}^{+\infty}S_{1}\left(  t,x\right)  \text{d}x}%
{\text{d}t}=  &  \lambda S_{0}^{d}\left(  t\right)  -\lambda\int_{0}^{+\infty
}S_{1}^{d}\left(  t,x\right)  \text{d}x\nonumber\\
&  -\int_{0}^{+\infty}\mu\left(  x\right)  S_{1}\left(  t,x\right)
\text{d}x+\int_{0}^{+\infty}\mu\left(  x\right)  S_{2}\left(  t,x\right)
\text{d}x, \label{Equ5}%
\end{align}
and for $k\geq2$,%
\begin{align}
\frac{\mathtt{d}\int_{0}^{+\infty}S_{k}\left(  t,x\right)  \text{d}x}%
{\text{d}t}=  &  \lambda\int_{0}^{+\infty}S_{k-1}^{d}\left(  t,x\right)
\text{d}x-\lambda\int_{0}^{+\infty}S_{k}^{d}\left(  t,x\right)  \text{d}%
x\nonumber\\
&  -\int_{0}^{+\infty}\mu\left(  x\right)  S_{k}\left(  t,x\right)
\text{d}x+\int_{0}^{+\infty}\mu\left(  x\right)  S_{k+1}\left(  t,x\right)
\text{d}x. \label{Equ6}%
\end{align}

\begin{Rem}
When there are $n$ servers in the supermarket model, it is necessary to give a
finite-size system of integral-differential equations for the fraction density
vector $S^{\left(  n\right)  }\left(  t,x\right)  =\left(  s_{0,n}\left(
t\right)  ,s_{1,n}\left(  t,x\right)  ,\ldots,s_{n,n}\left(  t,x\right)
\right)  $ as follows:%
\[
s_{0,n}\left(  t\right)  =1\text{ for all }t\geq0,
\]%
\[
\frac{\mathtt{d}}{\text{d}t}s_{0,n}\left(  t\right)  =-\lambda s_{0,n}%
^{d}\left(  t\right)  +\int_{0}^{+\infty}\mu\left(  x\right)  s_{1,n}\left(
t,x\right)  \text{d}x,
\]%
\begin{align*}
\frac{\mathtt{d}\int_{0}^{+\infty}s_{1,n}\left(  t,x\right)  \text{d}%
x}{\text{d}t}=  &  \lambda s_{0,n}^{d}\left(  t\right)  -\lambda\int
_{0}^{+\infty}s_{1,n}^{d}\left(  t,x\right)  \text{d}x\\
&  -\int_{0}^{+\infty}\mu\left(  x\right)  s_{1,n}\left(  t,x\right)
\text{d}x+\int_{0}^{+\infty}\mu\left(  x\right)  s_{2,n}\left(  t,x\right)
\text{d}x,
\end{align*}
and for $n\geq k\geq2$,%
\begin{align*}
\frac{\mathtt{d}\int_{0}^{+\infty}s_{k,n}\left(  t,x\right)  \text{d}%
x}{\text{d}t}=  &  \lambda\int_{0}^{+\infty}s_{k-1,n}^{d}\left(  t,x\right)
\text{d}x-\lambda\int_{0}^{+\infty}s_{k,n}^{d}\left(  t,x\right)  \text{d}x\\
&  -\int_{0}^{+\infty}\mu\left(  x\right)  s_{k,n}\left(  t,x\right)
\text{d}x+\int_{0}^{+\infty}\mu\left(  x\right)  s_{k+1,n}\left(  t,x\right)
\text{d}x.
\end{align*}
\end{Rem}

\section{Doubly Exponential Solution}

In this section, we discuss the fixed point of the system of
integral-differential equations in Equations (\ref{Equ3}) to (\ref{Equ6}), and
set up a system of nonlinear equations satisfied by the fixed point. Also, we
provide a closed-form solution: doubly exponential structure, to the system of
nonlinear equations.

A row vector $\pi\left(  x\right)  =\left(  \pi_{0},\pi_{1}\left(  x\right)
,\pi_{2}\left(  x\right)  ,\ldots\right)  $ is called a fixed point of the
fraction density vector $S\left(  t,x\right)  =\left(  S_{0}\left(  t\right)
,S_{1}\left(  t,x\right)  ,S_{2}\left(  t,x\right)  ,\ldots\right)  $ if there
exists a $t_{0}\geq0$ such that $S_{0}\left(  t\right)  =\pi_{0}$ and
$S_{k}\left(  t,x\right)  =\pi_{k}\left(  x\right)  $ for all $t\geq t_{0}$
and $k\geq1$. It is easy to see that if $\pi\left(  x\right)  $ is a fixed
point of the fraction density vector $S\left(  t,x\right)  $ for all $t\geq
t_{0}$, then%
\[
\frac{\mathtt{d}}{\text{d}t}S_{0}\left(  t\right)  _{|t\geq t_{0}}=0
\]
and for $k\geq1$%
\[
\frac{\mathtt{d}}{\text{d}t}S_{k}\left(  t,x\right)  _{|t\geq t_{0}}=0
\]
which leads to%
\begin{equation}
\int_{0}^{+\infty}\frac{\mathtt{d}}{\text{d}t}S_{k}\left(  t,x\right)
_{|t\geq t_{0}}\text{d}x=0. \label{Equ7}%
\end{equation}
Since for $k\geq1$%
\[
0\leq S_{k}\left(  t,x\right)  \leq1,
\]
using the Dominated Convergence Theorem we obtain%
\[
\frac{\mathtt{d}}{\text{d}t}\int_{0}^{+\infty}S_{k}\left(  t,x\right)
_{|t\geq t_{0}}\text{d}x=0.
\]
Therefore, if $\pi\left(  x\right)  =\left(  \pi_{0},\pi_{1}\left(  x\right)
,\pi_{2}\left(  x\right)  ,\ldots\right)  $ is a fixed point of the fraction
density vector $S\left(  t,x\right)  =\left(  S_{0}\left(  t\right)
,S_{1}\left(  t,x\right)  ,S_{2}\left(  t,x\right)  ,\ldots\right)  $ for all
$t\geq t_{0}$, then the system of integral-differential equations (\ref{Equ3})
to (\ref{Equ6}) can be simplified as%
\begin{equation}
\pi_{0}=1 \label{Equ8}%
\end{equation}%
\begin{equation}
-\lambda\pi_{0}^{d}+\int_{0}^{+\infty}\mu\left(  x\right)  \pi_{1}\left(
x\right)  \text{d}x=0, \label{Equ9}%
\end{equation}%
\begin{equation}
\lambda\pi_{0}^{d}\left(  t\right)  -\lambda\int_{0}^{+\infty}\pi_{1}%
^{d}\left(  x\right)  \text{d}x-\int_{0}^{+\infty}\mu\left(  x\right)  \pi
_{1}\left(  x\right)  \text{d}x+\int_{0}^{+\infty}\mu\left(  x\right)  \pi
_{2}\left(  x\right)  \text{d}x=0, \label{Equ10}%
\end{equation}
and for $k\geq2$,%
\begin{equation}
\lambda\int_{0}^{+\infty}\pi_{k-1}^{d}\left(  x\right)  \text{d}x-\lambda
\int_{0}^{+\infty}\pi_{k}^{d}\left(  x\right)  \text{d}x-\int_{0}^{+\infty}%
\mu\left(  x\right)  \pi_{k}\left(  x\right)  \text{d}x+\int_{0}^{+\infty}%
\mu\left(  x\right)  \pi_{k+1}\left(  x\right)  \text{d}x=0. \label{Equ11}%
\end{equation}

In what follows we derive a closed-form expression for $\pi\left(  x\right)
=\left(  \pi_{0},\pi_{1}\left(  x\right)  ,\pi_{2}\left(  x\right)
,\ldots\right)  $. It follows from Equations (\ref{Equ8}) and (\ref{Equ9})
that%
\begin{equation}
\int_{0}^{+\infty}\mu\left(  x\right)  \pi_{1}\left(  x\right)  \text{d}%
x=\lambda. \label{Equ12}%
\end{equation}
To solve Equation (\ref{Equ12}), using the fact that $\int_{0}^{+\infty}%
\mu\left(  x\right)  \overline{G}\left(  x\right)  $d$x=1$ we have%
\begin{equation}
\pi_{1}\left(  x\right)  =\lambda\overline{G}\left(  x\right)  =\rho\cdot
\mu\overline{G}\left(  x\right)  . \label{Equ13}%
\end{equation}
Based on the fact that $\pi_{0}=1$ and $\pi_{1}\left(  x\right)  =\rho\cdot
\mu\overline{G}\left(  x\right)  $, it follows from Equations (\ref{Equ10})
and (\ref{Equ12}) that%
\[
-\lambda\rho^{d}\cdot\int_{0}^{+\infty}\left[  \mu\overline{G}\left(
x\right)  \right]  ^{d}\text{d}x+\int_{0}^{+\infty}\mu\left(  x\right)
\pi_{2}\left(  x\right)  \text{d}x=0.
\]
Let $\theta=\int_{0}^{+\infty}\left[  \mu\overline{G}\left(  x\right)
\right]  ^{d}$d$x$, and we assume that $0<\theta<+\infty$. Then%
\begin{equation}
\int_{0}^{+\infty}\mu\left(  x\right)  \pi_{2}\left(  x\right)  \text{d}%
x=\lambda\theta\rho^{d}. \label{Equ14}%
\end{equation}
Using a similar analysis on Equation (\ref{Equ14}), we have%
\begin{equation}
\pi_{2}\left(  x\right)  =\lambda\theta\rho^{d}\overline{G}\left(  x\right)
=\theta\rho^{d+1}\cdot\mu\overline{G}\left(  x\right)  . \label{Equ15}%
\end{equation}
Based on $\pi_{1}\left(  x\right)  =\rho\cdot\mu\overline{G}\left(  x\right)
$ and $\pi_{2}\left(  x\right)  =\theta\rho^{d+1}\cdot\mu\overline{G}\left(
x\right)  $, we can compute%
\[
\lambda\int_{0}^{+\infty}\pi_{1}^{d}\left(  x\right)  \text{d}x=\lambda
\theta\rho^{d},
\]%
\[
\int_{0}^{+\infty}\mu\left(  x\right)  \pi_{2}\left(  x\right)  \text{d}%
x=\theta\rho^{d+1}\mu\int_{0}^{+\infty}\mu\left(  x\right)  \overline
{G}\left(  x\right)  \text{d}x=\lambda\theta\rho^{d}%
\]
and%
\[
\lambda\int_{0}^{+\infty}\pi_{2}^{d}\left(  x\right)  \text{d}x=\lambda
\theta^{d}\rho^{d^{2}+d}\int_{0}^{+\infty}\left[  \mu\overline{G}\left(
x\right)  \right]  ^{d}\text{d}x=\lambda\theta^{d+1}\rho^{d^{2}+d},
\]
thus it follows from Equation (\ref{Equ11}) that for $k=2$,%
\[
\int_{0}^{+\infty}\mu\left(  x\right)  \pi_{3}\left(  x\right)  \text{d}%
x=\lambda\theta^{d+1}\rho^{d^{2}+d},
\]
which leads to%
\begin{equation}
\pi_{3}\left(  x\right)  =\theta^{d+1}\rho^{d^{2}+d}\overline{G}\left(
x\right)  =\theta^{d+1}\rho^{d^{2}+d+1}\cdot\mu\overline{G}\left(  x\right)  .
\label{Equ15-1}%
\end{equation}

Based on the above analysis for the simple expressions $\pi_{k}\left(
x\right)  $ for $k=1,2$ and $3$, we can summarize the following theorem.

\begin{The}
\label{The:CFS}The fixed point $\pi=\left(  \pi_{0},\pi_{1}\left(  x\right)
,\pi_{2}\left(  x\right)  ,\ldots\right)  $ is given by%
\[
\pi_{0}=1,
\]%
\[
\pi_{1}\left(  x\right)  =\rho\cdot\mu\overline{G}\left(  x\right)
\]
and for $k\geq2,$%
\begin{equation}
\pi_{k}\left(  x\right)  =\theta^{d^{k-2}+d^{k-3}+\cdots+1}\rho^{d^{k-1}%
+d^{k-2}+\cdots+1}\cdot\mu\overline{G}\left(  x\right)  , \label{Equ16}%
\end{equation}
or%
\begin{equation}
\pi_{k}\left(  x\right)  =\theta^{\frac{d^{k-1}-1}{d-1}}\rho^{\frac{d^{k}%
-1}{d-1}}\cdot\mu\overline{G}\left(  x\right)  . \label{Equ17}%
\end{equation}
\end{The}

\noindent\textbf{Proof} By induction, one can easily derive the above result.

It is clear from (\ref{Equ15}) and (\ref{Equ15-1}) that Equation (\ref{Equ16})
or (\ref{Equ17}) is correct for the cases with $l=2,3$. Now, we assume that
Equation (\ref{Equ17}) is correct for the cases with $l=k$. Then%
\[
\lambda\int_{0}^{+\infty}\pi_{k-1}^{d}\left(  x\right)  \text{d}%
x=\lambda\theta^{\frac{d^{k-1}-1}{d-1}}\rho^{\frac{d^{k}-d}{d-1}},
\]%
\[
\int_{0}^{+\infty}\mu\left(  x\right)  \pi_{k}\left(  x\right)  \text{d}%
x=\lambda\theta^{\frac{d^{k-1}-1}{d-1}}\rho^{\frac{d^{k}-d}{d-1}}%
\]
and%
\[
\lambda\int_{0}^{+\infty}\pi_{k}^{d}\left(  x\right)  \text{d}x=\lambda
\theta^{\frac{d^{k}-1}{d-1}}\rho^{\frac{d^{k+1}-d}{d-1}},
\]
it follows from Equation (\ref{Equ11}) that%
\[
\int_{0}^{+\infty}\mu\left(  x\right)  \pi_{k+1}\left(  x\right)
\text{d}x=\lambda\theta^{\frac{d^{k}-1}{d-1}}\rho^{\frac{d^{k+1}-d}{d-1}}.
\]
Thus, for $l=k+1$ we have%
\[
\pi_{k+1}\left(  x\right)  =\theta^{\frac{d^{k}-1}{d-1}}\rho^{\frac{d^{k+1}%
-1}{d-1}}\cdot\mu\overline{G}\left(  x\right)  .
\]
This completes the proof. \hspace*{\fill} \rule{1.8mm}{2.5mm}

Let $\widetilde{\theta}=\int_{0}^{+\infty}\left[  \overline{G}\left(
x\right)  \right]  ^{d}$d$x$. Then $\theta=\mu^{d}\widetilde{\theta}$. The
following corollary provides another expression for the fixed point.

\begin{Cor}
\label{Cor:Decom}%
\[
\pi_{0}=1
\]
and for $k\geq1$%
\[
\pi_{k}\left(  x\right)  =\lambda^{\frac{d^{k}-1}{d-1}}\cdot\left\{
\widetilde{\theta}^{\frac{d^{k-1}-1}{d-1}}\overline{G}\left(  x\right)
\right\}  .
\]
\end{Cor}

It is easy to see from Corollary \ref{Cor:Decom} that the fixed point is
decomposited into two groups of information under a product form: the arrival
information and the service information. At the same time, the service
information indicates that the doubly exponential solution to the fixed point
must exist for $0<\mu<+\infty$, even if the service times are heavy-tailed.

The following corollary provides an upper bound for the fixed point.

\begin{Cor}
For $k\geq1$ and $x\geq0,$%
\[
\pi_{k}\left(  x\right)  <\int_{0}^{+\infty}\pi_{k}\left(  x\right)
\text{d}x<\rho^{\frac{d^{k-1}-1}{d-1}}\frac{\lambda^{d^{k}}}{\mu}.
\]
\end{Cor}

\textbf{Proof:} Note that $0\leq\overline{G}\left(  x\right)  \leq1$, we have%
\[
\widetilde{\theta}=\int_{0}^{+\infty}\left[  \overline{G}\left(  x\right)
\right]  ^{d}\text{d}x<\int_{0}^{+\infty}\overline{G}\left(  x\right)
\text{d}x=\frac{1}{\mu}.
\]
It follows from Corollary \ref{Cor:Decom} that%
\[
\pi_{k}\left(  x\right)  <\int_{0}^{+\infty}\pi_{k}\left(  x\right)
\text{d}x<\int_{0}^{+\infty}\lambda^{\frac{d^{k}-1}{d-1}}\widetilde{\theta
}^{\frac{d^{k-1}-1}{d-1}}\overline{G}\left(  x\right)  \text{d}x=\rho
^{\frac{d^{k-1}-1}{d-1}}\frac{\lambda^{d^{k}}}{\mu}.
\]
This completes the proof. \hspace*{\fill} \rule{1.8mm}{2.5mm}

Now, we compute the expected sojourn time $T_{d}$\ which a tagged arriving
customer spends in the supermarket model. For the general service times, a
tagged arriving customer is the $k$th customer in the corresponding queue with
the following probability%
\[
\int_{0}^{+\infty}\pi_{k-1}^{d}\left(  x\right)  \text{d}x-\int_{0}^{+\infty
}\pi_{k}^{d}\left(  x\right)  \text{d}x=\theta^{\frac{d^{k-2}-1}{d-1}}%
\rho^{\frac{d^{k-1}-1}{d-1}}-\theta^{\frac{d^{k-1}-1}{d-1}}\rho^{\frac{d^{k}%
-1}{d-1}}.
\]
When $k\geq1$, the head customer in the queue has been served, and so its
service time is residual and is denoted as $X_{R}$. Under the stationary
setting, we have%
\[
P\left\{  X_{R}\leq x\right\}  =\int_{0}^{x}\left[  \mu\overline{G}\left(
y\right)  \right]  \text{d}y
\]
with%
\[
E\left[  X_{R}\right]  =\int_{0}^{+\infty}\int_{x}^{+\infty}\left[
\mu\overline{G}\left(  y\right)  \right]  \text{d}y\text{d}x.
\]
Thus it is easy to see that the expected sojourn time of the tagged arriving
customer is given by%
\begin{align*}
E\left[  T_{d}\right]  =  &  \left[  \pi_{0}^{\odot d}-\int_{0}^{+\infty}%
\pi_{1}^{d}\left(  x\right)  \text{d}x\right]  E\left[  X\right] \\
&  +\sum_{k=1}^{\infty}\left[  \int_{0}^{+\infty}\pi_{k}^{d}\left(  x\right)
\text{d}x-\int_{0}^{+\infty}\pi_{k+1}^{d}\left(  x\right)  \text{d}x\right]
\left[  E\left[  X_{R}\right]  +kE\left[  X\right]  \right] \\
=  &  \left[  1-\int_{0}^{+\infty}\pi_{1}^{d}\left(  x\right)  \text{d}%
x\right]  E\left[  X\right]  +\int_{0}^{+\infty}\pi_{1}^{d}\left(  x\right)
\text{d}xE\left[  X_{R}\right] \\
&  +E\left[  X\right]  \sum_{k=1}^{\infty}k\left[  \int_{0}^{+\infty}\pi
_{k}^{d}\left(  x\right)  \text{d}x-\int_{0}^{+\infty}\pi_{k+1}^{d}\left(
x\right)  \text{d}x\right] \\
=  &  \left\{  E\left[  X_{R}\right]  -E\left[  X\right]  \right\}  \int
_{0}^{+\infty}\pi_{1}^{d}\left(  x\right)  \text{d}x+E\left[  X\right]
\left\{  1+\sum_{k=1}^{\infty}\int_{0}^{+\infty}\pi_{k}^{d}\left(  x\right)
\text{d}x\right\} \\
=  &  \theta\rho^{d}\left\{  E\left[  X_{R}\right]  -E\left[  X\right]
\right\}  +E\left[  X\right]  \left[  \sum_{k=1}^{\infty}\theta^{\frac{d^{k}%
-1}{d-1}}\rho^{\frac{d^{k}-d}{d-1}}\right]  .
\end{align*}

If the service times are exponential, then $E\left[  X_{R}\right]  =E\left[
X\right]  $, thus we obtain%
\[
E\left[  T_{d}\right]  =\frac{1}{\mu}\left[  \sum_{k=1}^{\infty}%
\theta^{\frac{d^{k}-1}{d-1}}\rho^{\frac{d^{k}-d}{d-1}}\right]  ,
\]
which is the same as Corollary 3.8 in Mitzenmacher \cite{Mit:1996b}.

We consider a computational example for the expected sojourn time in the
supermarket model with an Erlang service time distribution $E\left(
m,\mu\right)  $, where $m=2,\mu=1,d=2$. Figure 3 shows how the the expected
sojourn time depends on the arrival rate.

\begin{figure}[tbh]
\centering                       \includegraphics[width=8cm]{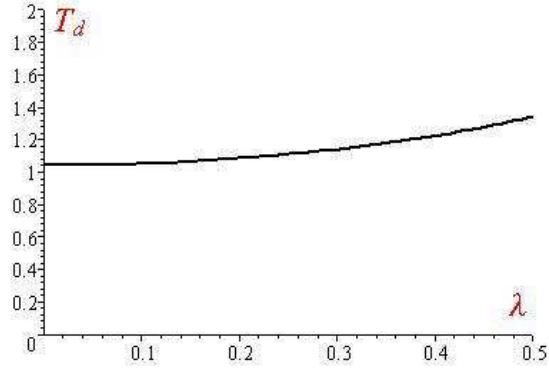}
\caption{the expected sojourn time $E\left[  T_{d}\right]  $ depends on the
arrival rate $\lambda$}%
\end{figure}

With the results from Equation (\ref{Equ16}) or (\ref{Equ17}), let us now
provide some useful discussions on the \emph{asymptotic behavior} of the fixed
point $\pi\left(  x\right)  =\left(  \pi_{0},\pi_{1}\left(  x\right)  ,\pi
_{2}\left(  x\right)  ,\ldots\right)  $. Note that we express $a_{k}\backsim
O\left(  b_{k}\right)  $ if $\lim_{k\rightarrow\infty}a_{k}/b_{k}=c\in\left(
-\infty,0\right)  \cup\left(  0,+\infty\right)  $. \vspace{0.1in}

\begin{Rem}
\noindent If the general distribution $G\left(  x\right)  $ and its mean
$1/\mu$ are given, then $\theta=\int_{0}^{+\infty}\left[  \mu\overline
{G}\left(  x\right)  \right]  ^{d}$d$x$ is a deterministic factor. We have%
\[
\frac{\pi_{k}\left(  x\right)  }{\theta^{\frac{d^{k-1}-1}{d-1}}}\sim O\left(
\rho^{\frac{d^{k}-1}{d-1}}\right)  \mu\overline{G}\left(  x\right)  ,\text{
\ as }k\rightarrow\infty.
\]
In this case, the heavy traffic should have a bigger influence on the
asymptotic behavior of the sequence $\left\{  \frac{\pi_{k}\left(  x\right)
}{\theta^{\frac{d^{k-1}-1}{d-1}}}\right\}  $.
\end{Rem}

\begin{Rem}
\vspace{0.2in} \noindent If $\rho$ is given, then%
\[
\frac{\pi_{k}\left(  x\right)  }{\rho^{\frac{d^{k}-1}{d-1}}}\sim O\left(
\theta^{\frac{d^{k-1}-1}{d-1}}\right)  \mu\overline{G}\left(  x\right)
,\text{ \ as }k\rightarrow\infty.
\]
In this case, the maximal value $\theta_{\max}$ of the positive number
$\int_{0}^{+\infty}\left[  \mu\overline{G}\left(  x\right)  \right]  ^{d}$d$x$
should have a bigger influence on the asymptotic behavior of the sequence
$\left\{  \frac{\pi_{k}\left(  x\right)  }{\rho^{\frac{d^{k}-1}{d-1}}%
}\right\}  $.
\end{Rem}

\section{A discussion for the key parameter $\theta$}

In this section, we provide a necessary discussion for the key parameter
$\theta$ in the doubly exponential solution of Theorem \ref{The:CFS}. Based on
this, for the fixed point we give a new and important observation: the doubly
exponential solution to the fixed point can extensively exist for
$0<\mu<+\infty$, even if the service time distribution is heavy-tailed.

Note that%
\begin{align}
\theta &  =\int_{0}^{+\infty}\left[  \mu\overline{G}\left(  x\right)  \right]
^{d}\text{d}x\nonumber\\
&  =\frac{\int_{0}^{+\infty}\left[  \overline{G}\left(  x\right)  \right]
^{d}\text{d}x}{\left[  \int_{0}^{+\infty}\overline{G}\left(  x\right)
\text{d}x\right]  ^{d}}, \label{Dequ1}%
\end{align}
it is easy to see that $\theta=1$ if $d=1$. Thus, we need to analyze the case
for $k\geq2$ as follows.

Since $0\leq\overline{G}\left(  x\right)  \leq1$, we get that $0\leq\left[
\overline{G}\left(  x\right)  \right]  ^{d}\leq\overline{G}\left(  x\right)
\leq1$, which leads to%
\[
\int_{0}^{+\infty}\left[  \overline{G}\left(  x\right)  \right]  ^{d}%
dx\leq\int_{0}^{+\infty}\overline{G}\left(  x\right)  dx=1/\mu.
\]
It is easy to see that $0<\theta<\mu^{d-1}$, and thus if $0<\mu<+\infty$, then
$0<\theta<+\infty$.

In what follows we analyze five simple and useful examples. In first two
examples, the service time distribution is light-tailed; while in the last
three examples, the service time distribution is heavy-tailed. Specifically,
the examples with heavy-tailed service times illustrate two important
observations: the first one indicates that the fixed point for the supermarket
model is different from the tail of stationary queue length distribution for
the ordinary M/G/1 queue, and the second one is to show that the doubly
exponential solution to the fixed point can exist extensively if the service
time mean is non-zero and finite.

\textbf{Example one: }Exponential distribution. Let $\overline{G}\left(
x\right)  =e^{-\mu x}$. Then $\theta=\mu^{d-1}/d$. It is easy to see that when
$\mu>\sqrt[d-1]{d}$, $\theta>1$; when $\mu=\sqrt[d-1]{d}$, $\theta=1$; and
when $\mu<\sqrt[d-1]{d}$, $0<\theta<1$. If $d=2$, then $\theta$ is a linear
function of $\mu$, and if $d=3$, then $\theta$ is a nonlinear function of
$\mu$. Figures 3 and 4 show the functions $\theta=\mu/2$ and $\theta=\mu
^{2}/3$, respectively.

\begin{figure}[tbh]
\centering             \includegraphics[width=8cm]{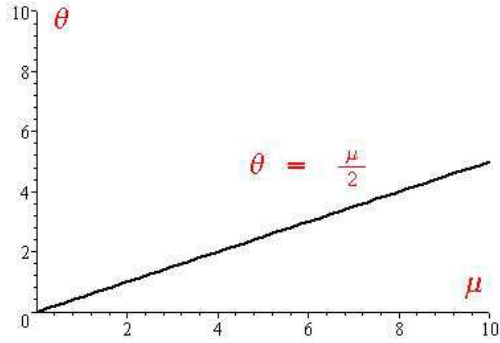} \caption{$\theta$
is a linear function of $\mu$ for $d=2$}%
\end{figure}

\begin{figure}[tbh]
\centering             \includegraphics[width=8cm]{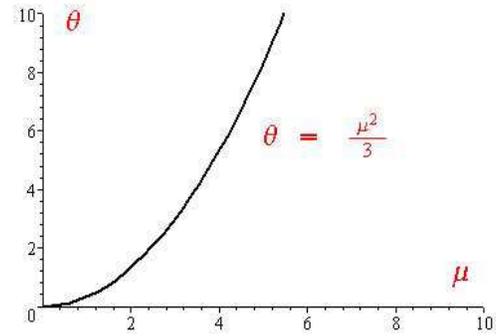} \caption{$\theta$
is a nonlinear function of $\mu$ for $d=3$}%
\end{figure}

\textbf{Example two: }Erlang distribution $E\left(  m,\mu\right)  $. Let
$\overline{G}\left(  x\right)  =e^{-\mu x}\sum_{k=0}^{m}\frac{\left(  \mu
x\right)  ^{k}}{k!}$. Then $\theta$ is given by%
\[
\theta=\left(  \frac{\mu}{m}\right)  ^{d}\int_{0}^{+\infty}e^{-\mu dx}\left[
\sum_{k=0}^{m}\frac{\left(  \mu x\right)  ^{k}}{k!}\right]  ^{d}\text{d}x.
\]
Let $\mu=1$. Table 1 lists how $\theta$ depends on the parameter pair $\left(
m,d\right)  $. As seen from Table 1, $\theta$ is decreasing for each of the
two parameters $m$ and $d$.

\begin{table}[tbh]
\caption{$\theta$ depends on the Erlang parameter pair $(m,d)$}%
\label{table: 1}
\centering
\begin{tabular}
[c]{|c||c|c|c|c|c||c|c|}\hline\hline
$(m, d)$ & (2, 2) & (2, 5) & (2, 10) & (5, 2) & (10, 2) & (5, 5) & (10,
10)\\\hline\hline
$\theta$ & 0.52 & 0.19 & $9.15\times10^{-2}$ & $4.13\times10^{-2}$ &
$9.48\times10^{-4}$ & $1.11\times10^{-3}$ & $6.51\times10^{-10}$\\\hline
\end{tabular}
\end{table}

\textbf{Example three:} Weibull distribution $W\left(  \tau,\mu\right)  $. Let
$\overline{G}\left(  x\right)  =\exp\left\{  -\left(  \mu t\right)  ^{\tau
}\right\}  $. It is easy to check that the mean of the Weibull distribution is
given by%
\[
\frac{1}{\mu}\Gamma\left(  1+\frac{1}{\tau}\right)  ,
\]
which follows that $\theta$ is given by%
\[
\theta=\frac{\mu^{d-1}}{d^{\frac{1}{\tau}}\left[  \Gamma\left(  1+\frac{1}%
{\tau}\right)  \right]  ^{d-1}},
\]
where $\Gamma(\alpha)=\int_{0}^{+\infty}x^{\alpha-1}e^{-x}$d$x$. Obviously,
the Weibull distribution $W\left(  \tau,\mu\right)  $ is heavy-tailed if
$0<\tau<1$; and the Weibull distribution $W\left(  \tau,\mu\right)  $ is
light-tailed if $\tau>1$. To indicate the role played by the heavy-tailed
parameter $\tau$ for $0<\tau<1$, taking $\mu=5$ and $d=2$ we have%
\[
\theta=\frac{5}{2^{\frac{1}{\tau}}\Gamma\left(  1+\frac{1}{\tau}\right)  }.
\]
Table 2 indicates how $\theta$ depends on the heavy-tailed parameter $\tau$,
such as, $0<\theta<1$ if $\tau=0.2$; $\theta>1$ if $\tau=0.9$. This example,
together with Theorem \ref{The:CFS}, illustrates an important observation that
the fixed point $\pi=\left(  \pi_{0},\pi_{1}\left(  x\right)  ,\pi_{2}\left(
x\right)  ,\ldots\right)  $ is doubly exponential (clearly, it is
light-tailed) even if the service time distribution is heavy-tailed. Based on
this, the the fixed point is different from the tail of stationary queue
length distribution of the ordinary M/G/1 queue, since for the ordinary M/G/1
queue, the stationary queue length distribution is heavy-tailed if the service
time distribution is heavy-tailed, e.g., see Adler, Feldman and Taqqu
\cite{Adl:1998}.

\begin{table}[tbh]
\caption{$\theta$ depends on the heavy-tailed parameter $\tau$ for $0<\tau<1$}%
\label{table: 2}
\centering
\begin{tabular}
[c]{|c||c|c|c|c|c||c|c|c|}\hline\hline
$\tau$ & 0.2 & 0.3 & 0.4 & 0.5 & 0.6 & 0.7 & 0.8 & 0.9\\\hline\hline
$\theta$ & $1.3\times10^{-3}$ & $5.3\times10^{-2}$ & 0.27 & 0.63 & 1.05 &
1.47 & 1.86 & 2.19\\\hline
\end{tabular}
\end{table}

\textbf{Example four:} Power law distribution. Let $\overline{G}\left(
x\right)  =\left(  \mu+x\right)  ^{-\alpha}$. If $0<\alpha\leq1$, then the
power law distribution does not exist the finite mean. In this case, we can
not setup the system of integral-differential equations for the fraction
density vector $S\left(  t,x\right)  =\left(  S_{0}\left(  t\right)
,S_{1}\left(  t,x\right)  ,S_{2}\left(  t,x\right)  ,\ldots\right)  $ which
leads to the analysis for the fixed point. Thus we only deal with the case
with $\alpha>1$. Note that for each $\alpha>1$
\[
\int_{0}^{+\infty}\overline{G}\left(  x\right)  \text{d}x=\frac{1}{\mu}%
\]
and%
\[
\int_{0}^{+\infty}\left[  \overline{G}\left(  x\right)  \right]  ^{d}%
\text{d}x=\frac{1}{\mu},
\]
thus we obtain $\theta=\mu^{d-1}$. It is easy to see that when $\mu>1$,
$\theta>1$; when $\mu=1$, $\theta=1$; and when $0<\mu<1$, $0<\theta<1$. It
follows from Theorem \ref{The:CFS} that for $k\geq1$%
\[
\pi_{k}\left(  x\right)  =\mu^{d^{k-1}-1}\rho^{\frac{d^{k}-1}{d-1}}\cdot
\mu\overline{G}\left(  x\right)  .
\]
This indicates that the fixed point $\pi=\left(  \pi_{0},\pi_{1}\left(
x\right)  ,\pi_{2}\left(  x\right)  ,\ldots\right)  $ is doubly exponential
(of course, it is light-tailed) if the service time distribution is power law.

\textbf{Example five: }Almost exponential distribution. Let $\overline
{G}\left(  x\right)  =\exp\left\{  -x\left(  \ln x\right)  ^{-\alpha}\right\}
$. Then it is easy to see that the almost exponential distribution is
heavy-tailed if $\alpha>0$
\[
\theta=\frac{\int_{0}^{+\infty}\exp\left\{  -dx\left(  \ln x\right)
^{-\alpha}\right\}  \text{d}x}{\left[  \int_{0}^{+\infty}\exp\left\{
-x\left(  \ln x\right)  ^{-\alpha}\right\}  \text{d}x\right]  ^{d}}%
\]
Table 3 lists how $\theta$ depends on the parameter pair $(d,\alpha)$. As seen
from Table 3, $\theta$ is decreasing for each of the two parameters $d$ and
$\alpha$.

\begin{table}[tbh]
\caption{$\theta$ depends on the parameter pair $(d, \alpha)$}%
\label{table: 3}
\centering
\begin{tabular}
[c]{|c||c|c|c|c|}\hline\hline
$(d, \alpha)$ & (2, 2) & (4, 2) & (2, 4) & (4, 4)\\\hline\hline
$\theta$ & $2.24\times10^{-2}$ & $2.01\times10^{-4}$ & $3.44\times10^{-5}$ &
$1.18\times10^{-13}$\\\hline
\end{tabular}
\end{table}

\section{The key parameter $\theta$ for PH Service Times}

In this section, as an important example we provide three methods to analyze a
supermarket model with Poisson arrivals and PH service times. Our purpose is
to provide three different ways to determine the key parameter $\theta$ and
compute the doubly exponential solution to the fixed point. Also, we indicate
that the doubly exponential solution to the fixed point is not unique for a
more general supermarket model.

The supermarket model with Poisson arrivals and PH service times is described
as follows. Customers arrive at a queueing system of $n>1$ servers as a
Poisson process with arrival rate $n\lambda$ for $\lambda>0$. The service
times of these customers are of phase type with irreducible representation
$\left(  \alpha,T\right)  $ of order $m$. Each arriving customer chooses
$d\geq1$ servers independently and uniformly at random from these $n$ servers,
and waits for service at the server which currently contains the fewest number
of customers. If there is a tie, servers with the fewest number of customers
will be chosen randomly. All customers in any service center will be served in
the FCFS manner. For the PH service time distribution, we use the irreducible
representation $\left(  \alpha,T\right)  $ of order $m$, where the row vector
$\alpha$ is a probability vector whose $j$th entry is the probability that a
service begins in phase $j$ for $1\leq j\leq m$; and $T$ is a matrix of order
$m$ whose $\left(  i,j\right)  ^{\text{th}}$ entry is denoted by $t_{i,j}$
with $t_{i,i}<0$ for $1\leq i\leq m$, and $t_{i,j}\geq0$ for $1\leq i,j\leq m$
and $i\neq j$. Let $T^{0}=-Te\gvertneqq0$, where $e$ is a column vector of
ones with a suitable dimension in the context. When a PH service time is in
phase $i$, the transition rate from phase $i$ to phase $j$ is $t_{i,j}$, the
service completion rate is $t_{i}^{0}$. At the same time, the mean service
rate is given by%
\[
\mu=-\frac{1}{\alpha T^{-1}e}.
\]
Unless we state otherwise, we assume that all the random variables defined
above are independent, and that the system is operating at the stable region:
$\rho=\lambda/\mu<1$.

We introduce some useful notation. Let $n_{k}^{\left(  i\right)  }\left(
t\right)  $ be the number of queues with at least $k$ customers and the
service time in phase $i$ at time $t\geq0$. Clearly, $0\leq n_{k}^{\left(
i\right)  }\left(  t\right)  \leq n$ for $1\leq i\leq m$ and $0\leq k\leq n$.
We define%
\[
s_{k}^{\left(  i\right)  }\left(  t\right)  =\frac{n_{k}^{\left(  i\right)
}\left(  t\right)  }{n},
\]
which is the fraction of queues with at least $k$ customers and the service
time in phase $i$. We write%
\[
S_{0}\left(  t\right)  =\left(  s_{0}\left(  t\right)  \right)
\]
and for $k\geq1$,%
\[
S_{k}\left(  t\right)  =\left(  s_{k}^{\left(  1\right)  }\left(  t\right)
,s_{k}^{\left(  2\right)  }\left(  t\right)  ,\ldots,s_{k}^{\left(  m\right)
}\left(  t\right)  \right)  ,
\]%
\[
S\left(  t\right)  =\left(  S_{0}\left(  t\right)  ,S_{1}\left(  t\right)
,S_{2}\left(  t\right)  ,\ldots\right)  .
\]
We now introduce Hadamard Product of two matrices $A=\left(  a_{i,j}\right)  $
and $B=\left(  b_{i,j}\right)  $ as follows:%
\[
A\odot B=\left(  a_{i,j}b_{i,j}\right)  .
\]
Specifically, for $k\geq2$ we have%
\[
A^{\odot k}=\underset{k\text{ matrix }A}{\underbrace{A\odot A\odot\cdots\odot
A}}.
\]
Let $a=\left(  a_{1},a_{2},a_{3},\ldots\right)  $. We write%
\[
a^{\odot\frac{1}{d}}=\left(  a_{1}^{\frac{1}{d}},a_{2}^{\frac{1}{d}}%
,a_{3}^{\frac{1}{d}},\ldots\right)  .
\]

Using a similar analysis to that in Equations (\ref{Equ3}) to (\ref{Equ6}), we
can obtain the following systems of differential vector equations for the
fraction density vector $S\left(  t\right)  =\left(  S_{0}\left(  t\right)
,S_{1}\left(  t\right)  ,S_{2}\left(  t\right)  ,\ldots\right)  $.%
\begin{equation}
S_{0}\left(  t\right)  =1,\text{ \ for }t\geq0, \label{Eq0}%
\end{equation}%
\begin{equation}
\frac{\mathtt{d}}{\text{d}t}S_{0}\left(  t\right)  =-\lambda S_{0}^{\odot
d}\left(  t\right)  +S_{1}\left(  t\right)  T^{0}, \label{Eq1}%
\end{equation}%
\begin{equation}
\frac{\mathtt{d}}{\text{d}t}S_{1}\left(  t\right)  =\lambda\alpha S_{0}^{\odot
d}\left(  t\right)  -\lambda S_{1}^{\odot d}\left(  t\right)  +S_{1}\left(
t\right)  T+S_{2}\left(  t\right)  T^{0}\alpha, \label{Eq2}%
\end{equation}
and for $k\geq2$,%
\begin{equation}
\frac{\mathtt{d}}{\text{d}t}S_{k}\left(  t\right)  =\lambda S_{k-1}^{\odot
d}\left(  t\right)  -\lambda S_{k}^{\odot d}\left(  t\right)  +S_{k}\left(
t\right)  T+S_{k+1}\left(  t\right)  T^{0}\alpha. \label{Eq3}%
\end{equation}

If $\pi=\left(  \pi_{0},\pi_{1},\pi_{2},\ldots\right)  $ is a fixed point of
the fraction density vector $S\left(  t\right)  $, then the system of
differential vector equations (\ref{Eq0}) to (\ref{Eq3}) can be simplified as%
\begin{equation}
\pi_{0}=1 \label{Eq3-1}%
\end{equation}%
\begin{equation}
-\lambda\pi_{0}^{\odot d}+\pi_{1}T^{0}=0, \label{Eq4}%
\end{equation}%
\begin{equation}
\lambda\alpha\pi_{0}^{\odot d}-\lambda\pi_{1}^{\odot d}+\pi_{1}T+\pi_{2}%
T^{0}\alpha=0, \label{Eq5}%
\end{equation}
and for $k\geq2$,%
\begin{equation}
\lambda\pi_{k-1}^{\odot d}-\lambda\pi_{k}^{\odot d}+\pi_{k}T+\pi_{k+1}%
T^{0}\alpha=0. \label{Eq6}%
\end{equation}

In what follows we provide three methods to solve the system of nonlinear
equations (\ref{Eq3-1}) to (\ref{Eq6}), and give three different doubly
exponential solutions to the fixed point.

\subsection{The first method}

The first method is based on Theorem \ref{The:CFS} given in this paper. For
the PH service time distribution%
\[
\overline{G}\left(  x\right)  =\alpha\exp\left\{  Tx\right\}  e
\]
Let $\theta=\int_{0}^{+\infty}\left[  \mu\overline{G}\left(  x\right)
\right]  ^{d}$d$x$, and we assume that $0<\theta<+\infty$. Then the fixed
point $\pi=\left(  \pi_{0},\pi_{1}\left(  x\right)  ,\pi_{2}\left(  x\right)
,\ldots\right)  $ is given by%
\[
\pi_{0}=1,
\]
and for $k\geq1$%
\begin{equation}
\pi_{k}\left(  x\right)  =\theta^{\frac{d^{k-1}-1}{d-1}}\rho^{\frac{d^{k}%
-1}{d-1}}\cdot\mu\overline{G}\left(  x\right)  . \label{Eq7}%
\end{equation}

\subsection{The second method}

The second method is proposed in Li, Wang and Liu \cite{Li:2010a}, and the key
parameter $\theta$\ is based on the stationary probability vector $\omega$ of
the irreducible Markov chain $T+T^{0}\alpha$, that is, $\theta=\omega^{\odot
d}e$.

It follows from Equation (\ref{Eq4}) that%
\[
\pi_{1}T^{0}=\lambda.
\]
Note that%
\[
\omega T^{0}=\mu,
\]%
\begin{equation}
\frac{\lambda}{\mu}\omega T^{0}=\lambda. \label{Eq8}%
\end{equation}
Thus, we obtain%
\[
\pi_{1}=\frac{\lambda}{\mu}\omega=\rho\cdot\,\omega.
\]
Based on the fact that $\pi_{0}=1$ and $\pi_{1}=\rho\cdot\omega$, it follows
from Equation (\ref{Eq5}) that%
\[
\lambda\alpha-\lambda\rho^{d}\cdot\omega^{\odot d}+\rho\cdot\omega T+\pi
_{2}T^{0}\alpha=0,
\]
which leads to%
\[
\lambda-\lambda\rho^{d}\cdot\omega^{\odot d}e+\rho\cdot\omega Te+\pi_{2}%
T^{0}=0.
\]
Note that $\omega Te=-\mu$ and $\rho=\lambda/\mu$, we obtain%
\[
\pi_{2}T^{0}=\lambda\rho^{d}\omega^{\odot d}e.
\]
Let $\theta=\omega^{\odot d}e$. Then it is easy to see that $\theta\in\left(
0,1\right)  $, and%
\[
\pi_{2}T^{0}=\lambda\theta\rho^{d}.
\]
Using a similar analysis on Equation (\ref{Eq8}), we have%
\[
\pi_{2}=\frac{\lambda\theta\rho^{d}}{\mu}\omega=\theta\rho^{d+1}\cdot\omega.
\]
Based on $\pi_{1}=\rho\,\omega$ and $\pi_{2}=\theta\rho^{d+1}\cdot\omega$, it
follows from Equation (\ref{Eq6}) that for $k=2$,%
\[
\lambda\rho^{d}\cdot\omega^{\odot d}-\lambda\theta^{d}\rho^{d^{2}+d}%
\cdot\omega^{\odot d}+\theta\rho^{d+1}\cdot\omega T+\pi_{3}T^{0}\alpha=0,
\]
which leads to%
\[
\lambda\theta\rho^{d}-\lambda\theta^{d+1}\rho^{d^{2}+d}+\theta\rho^{d+1}%
\cdot\omega Te+\pi_{3}T^{0}=0,
\]
thus we obtain%
\[
\pi_{3}T^{0}=\lambda\theta^{d+1}\rho^{d^{2}+d}.
\]
Using a similar analysis on Equation (\ref{Eq8}), we have%
\[
\pi_{3}=\frac{\lambda\theta^{d+1}\rho^{d^{2}+d}}{\mu}\omega=\theta^{d+1}%
\rho^{d^{2}+d+1}\cdot\omega.
\]
Now, we assume that $\pi_{k}=\theta^{\frac{d^{k-1}-1}{d-1}}\rho^{\frac{d^{k}%
-1}{d-1}}\cdot\omega$ is correct for the cases with $l=k$. Then it follows
from Equation (\ref{Eq6}) that for $l=k+1$, we have
\begin{align*}
\lambda &  \theta^{d^{k-2}+d^{k-3}+\cdots+d}\rho^{d^{k-1}+d^{k-2}+\cdots
+d}\cdot\omega^{\odot d}-\lambda\theta^{d^{k-1}+d^{k-2}+\cdots+d}\rho
^{d^{k}+d^{k-1}+\cdots+d}\cdot\omega^{\odot d}\\
&  +\theta^{d^{k-2}+d^{k-3}+\cdots+1}\rho^{d^{k-1}+d^{k-2}+\cdots+1}%
\cdot\omega T+\pi_{k+1}T^{0}\alpha=0,
\end{align*}
which leads to%
\begin{align*}
\lambda &  \theta^{d^{k-2}+d^{k-3}+\cdots+d+1}\rho^{d^{k-1}+d^{k-2}+\cdots
+d}-\lambda\theta^{d^{k-1}+d^{k-2}+\cdots+d+1}\rho^{d^{k}+d^{k-1}+\cdots+d}\\
&  +\theta^{d^{k-2}+d^{k-3}+\cdots+1}\rho^{d^{k-1}+d^{k-2}+\cdots+1}%
\cdot\omega Te+\pi_{k+1}T^{0}=0,
\end{align*}
thus we obtain%
\[
\pi_{k+1}T^{0}=\lambda\theta^{d^{k-1}+d^{k-2}+\cdots+d+1}\rho^{d^{k}%
+d^{k-1}+\cdots+d}.
\]
By a similar analysis to (\ref{Eq8}), we have%
\begin{align*}
\pi_{k+1}  &  =\frac{\lambda\theta^{d^{k-1}+d^{k-2}+\cdots+d+1}\rho
^{d^{k}+d^{k-1}+\cdots+d}}{\mu}\omega\\
&  =\theta^{d^{k-1}+d^{k-2}+\cdots+d+1}\rho^{d^{k}+d^{k-1}+\cdots+d+1}%
\cdot\omega.
\end{align*}
Therefore, by induction the fixed point $\pi=\left(  \pi_{0},\pi_{1},\pi
_{2},\ldots\right)  $ is given by%
\[
\pi_{0}=1,
\]
and for $k\geq1$%
\begin{equation}
\pi_{k}=\theta^{\frac{d^{k-1}-1}{d-1}}\rho^{\frac{d^{k}-1}{d-1}}\cdot\omega.
\label{Eq9}%
\end{equation}

\subsection{The third method}

The third method is based on the matrix computation for the system of
nonlinear equations (\ref{Eq3-1}) to (\ref{Eq6}), and shows that the key
parameter $\theta$ is based on the initial probability vector $\alpha$ in the
PH service time distribution, that is, $\theta=1/\alpha^{\odot\frac{1}{d}}e.$

It follows from (\ref{Eq3-1}) to (\ref{Eq6}) that
\begin{align*}
&  \left(  \pi_{1}^{\odot d},\pi_{2}^{\odot d},\pi_{3}^{\odot d}%
,\ldots\right)  \left(
\begin{array}
[c]{ccccc}%
-\lambda & \lambda &  &  & \\
& -\lambda & \lambda &  & \\
&  & -\lambda & \lambda & \\
&  &  & \ddots & \ddots
\end{array}
\right)  +\left(  \pi_{1},\pi_{2},\pi_{3},\ldots\right)  \left(
\begin{array}
[c]{cccc}%
T &  &  & \\
T^{0}\alpha & T &  & \\
& T^{0}\alpha & T & \\
&  & \ddots & \ddots
\end{array}
\right) \\
&  =-\left(  \lambda\alpha,0,0,\ldots\right)  ,
\end{align*}
which leads to%
\begin{align*}
\left(  \pi_{1},\pi_{2},\pi_{3},\ldots\right)  =  &  \left(  \pi_{1}^{\odot
d},\pi_{2}^{\odot d},\pi_{3}^{\odot d},\ldots\right)  \left(
\begin{array}
[c]{ccccc}%
R & V &  &  & \\
& R & V &  & \\
&  & R & V & \\
&  &  & \ddots & \ddots
\end{array}
\right) \\
&  +\left(  \lambda\alpha\left(  -T\right)  ^{-1},0,0,\ldots\right)  ,
\end{align*}
where%
\[
V=\lambda\left(  -T\right)  ^{-1}%
\]
and%
\[
R=\lambda\left(  -I+e\alpha\right)  \left(  -T\right)  ^{-1}.
\]
Thus we obtain%
\begin{equation}
\pi_{1}=\lambda\alpha\left(  -T\right)  ^{-1}+\pi_{1}^{\odot d}\left[
\lambda\left(  -I+e\alpha\right)  \left(  -T\right)  ^{-1}\right]
\label{Equ29}%
\end{equation}
and for $k\geq2$%
\begin{equation}
\pi_{k}=\pi_{k-1}^{\odot d}\left[  \lambda\left(  -T\right)  ^{-1}\right]
+\pi_{k}^{\odot d}\left[  \lambda\left(  -I+e\alpha\right)  \left(  -T\right)
^{-1}\right]  . \label{Equ30}%
\end{equation}
To omit the term $\pi_{k}^{\odot d}\left[  \lambda\left(  -I+e\alpha\right)
\left(  -T\right)  ^{-1}\right]  $ for $k\geq1$, we assume that $\left\{
\pi_{k},k\geq1\right\}  $ has the following expression%
\[
\pi_{k}=r\left(  k\right)  \alpha^{\odot\frac{1}{d}}.
\]
In this case, we have%
\[
\pi_{k}^{\odot d}\left[  \lambda\left(  -I+e\alpha\right)  \left(  -T\right)
^{-1}\right]  =r^{d}\left(  k\right)  \alpha\left[  \lambda\left(
-I+e\alpha\right)  \left(  -T\right)  ^{-1}\right]  =0,
\]
thus it follows from (\ref{Equ29}) and (\ref{Equ30}) that%
\begin{equation}
\pi_{1}=\lambda\alpha\left(  -T\right)  ^{-1} \label{Equ31}%
\end{equation}
and for $k\geq2$%
\begin{equation}
\pi_{k}=\pi_{k-1}^{\odot d}\left[  \lambda\left(  -T\right)  ^{-1}\right]  .
\label{Equ32}%
\end{equation}
It follows from (\ref{Equ31}) that%
\[
r\left(  1\right)  \alpha^{\odot\frac{1}{d}}=\lambda\alpha\left(  -T\right)
^{-1},
\]
which follows that%
\[
r\left(  1\right)  =\theta\rho,
\]
where%
\[
\theta=\frac{1}{\alpha^{\odot\frac{1}{d}}e}.
\]
It follows from (\ref{Equ32}) that%
\[
r\left(  k\right)  \alpha^{\odot\frac{1}{d}}=r^{d}\left(  k-1\right)
\alpha\left[  \lambda\left(  -T\right)  ^{-1}\right]  ,
\]
which follows that%
\[
r\left(  k\right)  =r^{d}\left(  k-1\right)  \theta\rho=\left(  \theta
\rho\right)  ^{\frac{d^{k}-1}{d-1}}.
\]
Therefore, we can obtain%
\[
\pi_{0}=1
\]
and for $k\geq1$%
\begin{equation}
\pi_{k}=\left(  \theta\rho\right)  ^{\frac{d^{k}-1}{d-1}}\cdot\alpha
^{\odot\frac{1}{d}}. \label{Equ33}%
\end{equation}

\subsection{Non-uniqueness}

Based on the above three methods, we can summarize the key parameter and the
doubly exponential solution to the fixed point in the following table.

\begin{table}[tbh]
\caption{Comparison for the three methods}%
\label{table: three methods}%
\centering
\begin{tabular}
[c]{|c||c|c|}\hline
& Key Parameter & Fixed point\\\hline\hline
Method 1 & $\theta=\frac{\int_{0}^{+\infty}\left[  \alpha\exp\left\{
Tx\right\}  e\right]  ^{d}\text{d}x}{\left[  -\alpha T^{-1}e\right]  ^{d}}$ &
$\pi_{k}=\theta^{\frac{d^{k-1}-1}{d-1}}\rho^{\frac{d^{k} -1}{d-1}}\cdot
\mu\overline{G}\left(  x\right)  $\\\hline
Method 2 & $\theta=\omega^{\odot d}e$ & $\pi_{k}=\theta^{\frac{d^{k-1}-1}%
{d-1}}\rho^{\frac{d^{k}-1}{d-1}}\cdot\omega$\\\hline
Method 3 & $\theta=1/\alpha^{\odot\frac{1}{d}}e$ & $\pi_{k}=\left(  \theta
\rho\right)  ^{\frac{d^{k}-1}{d-1}}\cdot\alpha^{\odot\frac{1}{d}}$\\\hline
\end{tabular}
\end{table}

When the PH service time is an $m$-order Erlang distribution with the
irreducible representation $(\alpha,T)$, where%
\[
\alpha=\left(  1,0,\ldots,0\right)
\]
and%
\[
T=\left(
\begin{array}
[c]{ccccc}%
-\eta & \eta &  &  & \\
& -\eta & \eta &  & \\
&  & \ddots & \ddots & \\
&  &  & -\eta & \eta\\
&  &  &  & -\eta
\end{array}
\right)  ,\text{ \ \ }T^{0}=\left(
\begin{array}
[c]{c}%
0\\
0\\
\vdots\\
0\\
\eta
\end{array}
\right)  .
\]
We have%
\[
\alpha^{\odot\frac{1}{d}}=\left(  1,0,\ldots,0\right)
\]
and%
\[
\theta=\frac{1}{\alpha^{\odot\frac{1}{d}}e}=1.
\]
Thus the doubly exponential solution by the third method is given by%
\begin{equation}
\pi_{k}=\rho^{\frac{d^{k}-1}{d-1}}\cdot\left(  1,0,\ldots,0\right)  ,\text{
\ }k\geq1. \label{Equ37}%
\end{equation}
It is clear that%
\[
T+T^{0}\alpha=\left(
\begin{array}
[c]{ccccc}%
-\eta & \eta &  &  & \\
& -\eta & \eta &  & \\
&  & \ddots & \ddots & \\
&  &  & -\eta & \eta\\
\eta &  &  &  & -\eta
\end{array}
\right)  ,
\]
which leads to the stationary probability vector of the Markov chain
$T+T^{0}\alpha$ as follows:
\[
\omega=\left(  \frac{1}{m},\frac{1}{m},\ldots,\frac{1}{m}\right)  ,
\]%
\[
\mu=\omega T^{0}=\frac{\eta}{m},
\]%
\[
\rho=\frac{\lambda}{\mu}=\frac{m\lambda}{\eta}%
\]
and%
\[
\theta=\omega^{\odot d}e=m\left(  \frac{1}{m}\right)  ^{d}=m^{1-d}.
\]
Thus the doubly exponential solution by the second method is given by
\begin{align}
\pi_{k}  &  =\theta^{\frac{d^{k-1}-1}{d-1}}\rho^{\frac{d^{k}-1}{d-1}}\left(
\frac{1}{m},\frac{1}{m},\ldots,\frac{1}{m}\right) \nonumber\\
&  =\rho^{\frac{d^{k}-1}{d-1}}\cdot\left(  m^{-d^{k-1}},m^{-d^{k-1}}%
,\ldots,m^{-d^{k-1}}\right)  ,\text{ \ }k\geq1. \label{Equ38}%
\end{align}
It is clear that the three doubly exponential solutions (\ref{Eq7}),
(\ref{Equ37}) and (\ref{Equ38}) are different for $m\geq2$.

\begin{Rem}
For the supermarket model with Poisson arrivals and PH service times, we have
obtained three different doubly exponential solutions to the fixed point. It
is interesting but difficult how to be able to find another new doubly
exponential solution. We believe that it is an open problem how to give all
the doubly exponential solutions to the fixed point for a more general
supermarket model including the case with MAP arrivals, PH service times or
general service times.
\end{Rem}

\section{Exponential convergence to the fixed point}

In this section, we study the exponential convergence of the current location
$S\left(  t,x\right)  $ of the supermarket model to its fixed point
$\pi\left(  x\right)  $ for $t\geq0$ and $x\geq0$. Not only does the
exponential convergence indicates the existence of the fixed point, but it
also explains such a convergent process is very fast.

For the supermarket model, the initial point $S\left(  0,x\right)  $ can
affect the current location $S\left(  t,x\right)  $ for each $t>0$, since the
service process in the supermarket model is under a unified structure. Here,
we provide notation for comparison of two vectors. Let $a=\left(  a_{1}%
,a_{2},a_{3},\ldots\right)  $ and $b=\left(  b_{1},b_{2},b_{3},\ldots\right)
$. We write $a\prec b$ if $a_{k}<b_{k}$ for all $k\geq1$; $a\preceq b$ if
$a_{k}\leq b_{k}$ for all $k\geq1$.

Now, we can obtain the following useful proposition whose proof is clear from
a sample path analysis and is omitted here.

\begin{Pro}
\label{Prop1}If $S\left(  0,x\right)  \preceq\widetilde{S}\left(  0,x\right)
$, then $S\left(  t,x\right)  \preceq\widetilde{S}\left(  t,x\right)  $.
\end{Pro}

Based on Proposition \ref{Prop1}, the following theorem shows that the fixed
point $\pi\left(  x\right)  $ is an upper bound of the current location
$S\left(  t,x\right)  $ for all $t\geq0$ and $x\geq0$.

\begin{The}
\label{The:UpperB}For the supermarket model, if there exists some $k$ such
that $S_{k}\left(  0,x\right)  =0$, then the sequence $\left\{  S_{k}\left(
t,x\right)  \right\}  $ has a upper bound sequence which decreases doubly
exponentially for all $t\geq0$ and $x\geq0$, that is, $S\left(  t,x\right)
\preceq\pi\left(  x\right)  $ for all $t\geq0$ and $x\geq0$.
\end{The}

\noindent\textbf{Proof} \ Let%
\[
\widetilde{S}_{0}\left(  0\right)  =\pi_{0}%
\]%
\[
\widetilde{S}_{k}\left(  0,x\right)  =\pi_{k}\left(  x\right)  ,\text{
\ }k\geq1.
\]
Then $\widetilde{S}_{0}\left(  t\right)  =\widetilde{S}_{0}\left(  0\right)
=\pi_{0}$, and for each $k\geq1$, $\widetilde{S}_{k}\left(  t,x\right)
=\widetilde{S}_{k}\left(  0,x\right)  =\pi_{k}\left(  x\right)  $ for all
$t\geq0$ and $x\geq0$, since $\widetilde{S}\left(  0\right)  $ is a fixed
point in the supermarket model. If $S_{k}\left(  0,x\right)  =0$ for some $k$,
then $S_{k}\left(  0,x\right)  <\widetilde{S}_{k}\left(  0,x\right)  $ and
$S_{j}\left(  0,x\right)  \leq\widetilde{S}_{j}\left(  0,x\right)  $\ for all
$j\geq1$ and $j\neq k$, thus $S\left(  0,x\right)  \preceq\widetilde{S}\left(
0,x\right)  $. It is easy to see from Proposition \ref{Prop1} that
$S_{k}\left(  t,x\right)  \leq\widetilde{S}_{k}\left(  t,x\right)  =\pi
_{k}\left(  x\right)  $ for all $k\geq1$, $t\geq0$ and $x\geq0$. Thus we
obtain that for all $k\geq1$, $t\geq0$ and $x\geq0$
\[
S_{k}\left(  t,x\right)  \leq\pi_{k}\left(  x\right)  =\theta^{\frac{d^{k-1}%
-1}{d-1}}\rho^{\frac{d^{k}-1}{d-1}}\cdot\left[  \mu\overline{G}\left(
x\right)  \right]  .
\]
This completes the proof. \nopagebreak                  \hspace*{\fill}
\nopagebreak        \textbf{{\rule{0.2cm}{0.2cm}}} \vspace{2.5ex}

To show the exponential convergence, we use Theorem \ref{The:UpperB} to define
a potential function (or Lyapunov function) $\Phi\left(  t\right)  $ as
follows:%
\[
\Phi\left(  t\right)  =\sum_{k=1}^{\infty}w_{k}\int_{0}^{+\infty}\left[
\pi_{k}\left(  x\right)  -S_{k}\left(  t,x\right)  \right]  \text{d}x,
\]
where $\left\{  w_{k}\right\}  $ is a positive scalar sequence with
$w_{k}>w_{k-1}\geq w_{1}=1$ for $k\geq2$. Note that $\pi_{0}=S_{0}\left(
t\right)  =1$. It is easy to see from Proposition \ref{Prop1} that
$\Phi\left(  t\right)  \geq0$ for all $t\geq0$.

When $\int_{0}^{+\infty}\left[  \pi_{k}\left(  x\right)  -S_{k}\left(
t,x\right)  \right]  $d$x>0$ for $k\geq1$, we write%
\[
\frac{\int_{0}^{+\infty}S_{k}^{d}\left(  t,x\right)  \text{d}x}{\int
_{0}^{+\infty}\left[  \pi_{k}\left(  x\right)  -S_{k}\left(  t,x\right)
\right]  \text{d}x}=c_{k}\left(  t\right)
\]
and%
\[
\frac{\int_{0}^{+\infty}\mu\left(  x\right)  S_{k}\left(  t,x\right)
\text{d}x}{\int_{0}^{+\infty}\left[  \pi_{k}\left(  x\right)  -S_{k}\left(
t,x\right)  \right]  \text{d}x}=d_{k}\left(  t\right)  .
\]

The following lemma provide a method to determine the positive scalar sequence
$\left\{  w_{k}\right\}  $ with $w_{k}>w_{k-1}\geq w_{1}=1$ for $k\geq2$. This
proof is easy by means of some simple computation.

\begin{Lem}
\label{Lem:PositiveC}If $\delta$ is a positive constant,
\[
w_{1}=1,
\]%
\[
\lambda\left(  w_{1}-w_{2}\right)  c_{1}\left(  t\right)  =-\delta w_{1}%
\]
and for $k\geq2$%
\[
\lambda\left(  w_{k}-w_{k+1}\right)  c_{k}\left(  t\right)  +\left(
w_{k}-w_{k-1}\right)  d_{k}\left(  t\right)  =-\delta w_{k},
\]
then%
\[
w_{2}=1+\frac{\delta}{\lambda c_{1}\left(  t\right)  },
\]
and for $k\geq3$%
\[
w_{k}=w_{k-1}+\frac{\delta w_{k-1}+\left(  w_{k-1}-w_{k-2}\right)
d_{k-1}\left(  t\right)  }{\lambda c_{k-1}\left(  t\right)  }.
\]
\end{Lem}

The following theorem measures the distance $\Phi\left(  t\right)  $ of the
current location $S\left(  t,x\right)  $ for $t\geq0$ and the fixed point
$\pi\left(  x\right)  $ for $x\geq0$, and illustrates that the distance
$\Phi\left(  t\right)  $ to the fixed point from the current location is very
go to zero with exponential convergence. Hence, it shows that from any
suitable starting point, the supermarket model can be quickly close to the
fixed point, that is, there always exists a fixed point in the supermarket model.

\begin{The}
For $t\geq0$,%
\[
\Phi\left(  t\right)  \leq c_{0}e^{-\delta t},
\]
where $c_{0}$ and $\delta$ are two positive constants, and they possibly
depend on time $t\geq0$. In this case, the potential function $\Phi\left(
t\right)  $ is exponentially convergent.
\end{The}

\noindent\textbf{Proof} \ Note that%
\[
\Phi\left(  t\right)  =\sum_{k=1}^{\infty}w_{k}\int_{0}^{+\infty}\left[
\pi_{k}\left(  x\right)  -S_{k}\left(  t,x\right)  \right]  \text{d}x,
\]
we have%
\begin{align*}
\frac{\text{d}}{\text{d}t}\Phi\left(  t\right)   &  =\frac{\text{d}}%
{\text{d}t}\sum_{k=1}^{\infty}w_{k}\int_{0}^{+\infty}\left[  \pi_{k}\left(
x\right)  -S_{k}\left(  t,x\right)  \right]  \text{d}x\\
&  =-\sum_{k=1}^{\infty}w_{k}\frac{\text{d}}{\text{d}t}\int_{0}^{+\infty}%
S_{k}\left(  t,x\right)  \text{d}x
\end{align*}
by means of the Dominated Convergence Theorem. It follows from (\ref{Equ3}) to
(\ref{Equ6}) that%
\begin{equation}
\int_{0}^{+\infty}\mu\left(  x\right)  S_{1}\left(  t,x\right)  \text{d}%
x=\lambda, \label{Eq4-1}%
\end{equation}
and using (\ref{Eq4-1}) we obtain%
\begin{align*}
\frac{d}{dt}\Phi\left(  t\right)  =  &  -\sum_{k=1}^{\infty}w_{k}%
\frac{\mathtt{d}}{\text{d}t}\int_{0}^{+\infty}S_{k}\left(  t,x\right)
\text{d}x\\
=  &  -w_{1}[\lambda-\lambda\int_{0}^{+\infty}S_{1}^{d}\left(  t,x\right)
\text{d}x\\
&  -\int_{0}^{+\infty}\mu\left(  x\right)  S_{1}\left(  t,x\right)
\text{d}x+\int_{0}^{+\infty}\mu\left(  x\right)  S_{2}\left(  t,x\right)
\text{d}x]\\
&  -\sum_{k=2}^{\infty}w_{k}[\lambda\int_{0}^{+\infty}S_{k-1}^{d}\left(
t,x\right)  \text{d}x-\lambda\int_{0}^{+\infty}S_{k}^{d}\left(  t,x\right)
\text{d}x\\
&  -\int_{0}^{+\infty}\mu\left(  x\right)  S_{k}\left(  t,x\right)
\text{d}x+\int_{0}^{+\infty}\mu\left(  x\right)  S_{k+1}\left(  t,x\right)
\text{d}x]\\
=  &  -w_{1}[-\lambda\int_{0}^{+\infty}S_{1}^{d}\left(  t,x\right)
\text{d}x+\int_{0}^{+\infty}\mu\left(  x\right)  S_{2}\left(  t,x\right)
\text{d}x]\\
&  -\sum_{k=2}^{\infty}w_{k}[\lambda\int_{0}^{+\infty}S_{k-1}^{d}\left(
t,x\right)  \text{d}x-\lambda\int_{0}^{+\infty}S_{k}^{d}\left(  t,x\right)
\text{d}x\\
&  -\int_{0}^{+\infty}\mu\left(  x\right)  S_{k}\left(  t,x\right)
\text{d}x+\int_{0}^{+\infty}\mu\left(  x\right)  S_{k+1}\left(  t,x\right)
\text{d}x],
\end{align*}
which follows
\begin{align*}
\frac{d}{dt}\Phi\left(  t\right)  =  &  \lambda\left(  w_{1}-w_{2}\right)
c_{1}\left(  t\right)  \int_{0}^{+\infty}\left[  \pi_{k}\left(  x\right)
-S_{k}\left(  t,x\right)  \right]  \text{d}x\\
&  +\sum_{k=2}^{\infty}\left[  \lambda\left(  w_{k}-w_{k+1}\right)
c_{k}\left(  t\right)  +\left(  w_{k}-w_{k-1}\right)  d_{k}\left(  t\right)
\right] \\
&  \cdot\int_{0}^{+\infty}\left[  \pi_{k}\left(  x\right)  -S_{k}\left(
t,x\right)  \right]  \text{d}x.
\end{align*}
Using Lemma \ref{Lem:PositiveC} we can easily choose a parameter $\delta>0$
and a suitable positive scalar sequence $\left\{  w_{k}\right\}  $ with
$w_{k}>w_{k-1}\geq w_{1}=1$ for $k\geq2$ such that%
\[
\lambda\left(  w_{1}-w_{2}\right)  c_{1}\left(  t\right)  \leq-\delta w_{1}%
\]
and for $k\geq2$%
\[
\left(  w_{k}-w_{k-1}\right)  d_{k}\left(  t\right)  -\lambda\left(
w_{k+1}-w_{k}\right)  c_{k}\left(  t\right)  \leq-\delta w_{k},
\]
thus we can obtain%
\[
\frac{d}{dt}\Phi\left(  t\right)  \leq-\delta\Phi\left(  t\right)  ,
\]
which leads to
\[
\Phi\left(  t\right)  \leq c_{0}e^{-\delta t}.
\]
This completes the proof. \nopagebreak    \hspace*{\fill} \nopagebreak
\textbf{{\rule{0.2cm}{0.2cm}}}

\begin{Rem}
We have provided an algorithm for computing the positive scalar sequence
$\left\{  w_{k}\right\}  $ with $1=w_{1}\leq w_{k-1}<w_{k}$ for $k\geq2$ as follows:

Step one:%
\[
w_{1}=1.
\]

Step two:%
\[
w_{2}=1+\frac{\delta}{\lambda c_{1}\left(  t\right)  }.
\]

Step three: for $k\geq2$%
\[
w_{k}=w_{k-1}+\frac{\delta w_{k-1}+\left(  w_{k-1}-w_{k-2}\right)
d_{k-1}\left(  t\right)  }{\lambda c_{k-1}\left(  t\right)  }.
\]
This illustrates that $w_{k}$ is a function of time $t$. Note that
$\lambda,\delta,c_{k}\left(  t\right)  ,d_{l}\left(  t\right)  >0$, it is
clear that for $k\geq2$%
\[
1=w_{1}\leq w_{k-1}<w_{k}.
\]
\end{Rem}

\section{Lipschitz Condition}

In this section, we apply the Kurtz Theorem to study the supermarket model
with general service times, and analyze the Lipschitz condition with respect
to general service times.

The supermarket model can be analyzed by a density dependent jump Markov
process, where the density dependent jump Markov process is a Markov process
with a single parameter $n$ which corresponds to the population size. Kurtz's
work provides a basis for the density dependent jump Markov processes in order
to relate the infinite-size system of differential equations to the
corresponding finite-size system of differential equations. Readers may refer
to Kurtz \cite{Kur:1981} for more details.

In the supermarket model, the states of density dependent jump Markov process
can be normalized and interpreted as measuring population densities, so that
the transition rates depend only on these densities. Hence, the infinite-size
system of differential equations can be regarded as the limiting model of the
corresponding finite-size system of differential equations as the population
size grows arbitrarily large. When the population size is $n$, we write%
\[
E_{n}=\left\{  k:k=0,1,\ldots,n\right\}  .
\]
For $k\geq1$ and $x\geq0$, we write%
\[
s_{k}^{\left(  n\right)  }\left(  x\right)  =\left(  \frac{k}{n},x\right)  ,
\]
where $x$ is the residual service time of each server, and%
\[
s_{k}^{\left(  n\right)  }=\int_{0}^{+\infty}s_{k}^{\left(  n\right)  }\left(
x\right)  \text{d}x.
\]
Let%
\[
S_{0}=\lim_{n\rightarrow\infty}s_{0}^{\left(  n\right)  }%
\]
and for $k\geq1$%
\[
S_{k}=\lim_{n\rightarrow\infty}\int_{0}^{+\infty}s_{k}^{\left(  n\right)
}\left(  x\right)  \text{d}x.
\]

Let $\left\{  \widehat{X}_{n}\left(  t\right)  :t\geq0\right\}  $ be a density
dependent jump Markov process on the state space $E_{n}$ whose transition
rates are given by%
\[
q_{k,k+l}^{\left(  n\right)  }=n\beta_{l}\left(  \frac{k}{n}\right)
=n\beta_{l}\left(  s_{k}^{\left(  n\right)  }\right)  .
\]
In this supermarket model, $\widehat{X}_{n}\left(  t\right)  $ is the unscaled
process which records the number of servers with at least $k$ customers for
$0\leq k\leq n$.

Let $a$ and $b$ denote an arrival and a service completion, respectively.
Hence taking $l=a$ or $b$ for $a>b>0$, we write
\[
\beta_{a}\left(  s_{0}^{\left(  n\right)  }\right)  =-\lambda,
\]%
\[
\beta_{b}\left(  s_{0}^{\left(  n\right)  }\right)  =\int_{0}^{+\infty}%
\mu\left(  x\right)  s_{1}^{\left(  n\right)  }\left(  x\right)  \text{d}x;
\]%
\[
\beta_{a}\left(  s_{1}^{\left(  n\right)  }\right)  =\lambda-\lambda\int
_{0}^{+\infty}\left[  s_{1}^{\left(  n\right)  }\left(  x\right)  \right]
^{d}\text{d}x,
\]%
\[
\beta_{b}\left(  s_{1}^{\left(  n\right)  }\right)  =-\int_{0}^{+\infty}%
\mu\left(  x\right)  s_{1}^{\left(  n\right)  }\left(  x\right)
\text{d}x+\int_{0}^{+\infty}\mu\left(  x\right)  s_{2}^{\left(  n\right)
}\left(  x\right)  \text{d}x;
\]
and for $n\geq k\geq2,$%
\[
\beta_{a}\left(  s_{k}^{\left(  n\right)  }\right)  =\lambda\int_{0}^{+\infty
}\left[  s_{k-1}^{\left(  n\right)  }\left(  x\right)  \right]  ^{d}%
\text{d}x-\lambda\int_{0}^{+\infty}\left[  s_{k}^{\left(  n\right)  }\left(
x\right)  \right]  ^{d}\text{d}x,
\]%
\[
\beta_{b}\left(  s_{k}^{\left(  n\right)  }\right)  =-\int_{0}^{+\infty}%
\mu\left(  x\right)  s_{k}^{\left(  n\right)  }\left(  x\right)
\text{d}x+\int_{0}^{+\infty}\mu\left(  x\right)  s_{k+1}^{\left(  n\right)
}\left(  x\right)  \text{d}x.
\]

Using Chapter 7 in Kurtz \cite{Kur:1981} or Subsection 3.4.1 in Mitzenmacher
\cite{Mit:1996b}, the Markov process $\left\{  \widehat{X}_{n}\left(
t\right)  :t\geq0\right\}  $ with transition rates $q_{k,k+l}^{\left(
n\right)  }$ is given by%
\begin{equation}
\widehat{X}_{n}\left(  t\right)  =\widehat{X}_{n}\left(  0\right)
+\sum_{l=a,b}lY_{l}\left(  n\int_{0}^{t}\beta_{l}\left(  \frac{\widehat{X}%
_{n}\left(  u\right)  }{n}\right)  \text{d}u\right)  , \label{Eq7.1}%
\end{equation}
where $Y_{l}\left(  x\right)  $ for $l=a$ and $b$ are two independent standard
Poisson processes. Clearly, the jump Markov process by Equation (\ref{Eq7.1})
at time $t$ is determined by the starting point and the transition rates which
are integrated over its history.

Let%
\begin{equation}
F\left(  y\right)  =a\beta_{a}\left(  y\right)  +b\beta_{b}\left(  y\right)  .
\label{Equ7.4}%
\end{equation}
Taking $X_{n}\left(  t\right)  =n^{-1}\widehat{X}_{n}\left(  t\right)  $ which
is an appropriate scaled process, we have%
\begin{equation}
X_{n}\left(  t\right)  =X_{n}\left(  0\right)  +\sum_{l=a,b}ln^{-1}\widehat
{Y}_{l}\left(  n\int_{0}^{t}\beta_{l}\left(  X_{n}\left(  u\right)  \right)
\text{d}u\right)  +\int_{0}^{t}F\left(  X_{n}\left(  u\right)  \right)
\text{d}u, \label{Equ7.5}%
\end{equation}
where $\widehat{Y}_{l}\left(  y\right)  =Y_{l}\left(  y\right)  -y$ is a
Poisson process centered at its expectation. Note that in (\ref{Equ7.5}), the
function $F\left(  y\right)  $ given in (\ref{Equ7.4}) is for $y=s_{k}%
^{\left(  n\right)  },0\leq k\leq n$.

Taking $X\left(  t\right)  =\lim_{n\rightarrow\infty}X_{n}\left(  t\right)  $
and $x_{0}=\lim_{n\rightarrow\infty}X_{n}\left(  0\right)  $, we obtain%
\begin{equation}
X\left(  t\right)  =x_{0}+\int_{0}^{t}F\left(  X\left(  u\right)  \right)
\text{d}u,\text{ \ }t\geq0, \label{Equ7.6}%
\end{equation}
due to the fact that%
\[
\lim_{n\rightarrow\infty}\frac{1}{n}\widehat{Y}_{l}\left(  n\int_{0}^{t}%
\beta_{l}\left(  X_{n}\left(  u\right)  \right)  \text{d}u\right)  =0
\]
by means of the law of large numbers. Note that in (\ref{Equ7.6}), the
function $F\left(  y\right)  $ given in (\ref{Equ7.4}) is for $y=S_{k},k\geq
1$. In the supermarket model, the deterministic and continuous process
$\left\{  X\left(  t\right)  ,t\geq0\right\}  $ is described by the
infinite-size system of integral-differential equations (\ref{Equ3}) to
(\ref{Equ6}), or simply in the below%
\begin{equation}
\frac{d}{dt}X\left(  t\right)  =F\left(  X\left(  t\right)  \right)
\label{Equ7.2}%
\end{equation}
with the initial condition%
\begin{equation}
X\left(  0\right)  =x_{0}. \label{Equ7.3}%
\end{equation}

Now, we consider the uniqueness of the limiting deterministic process
$\left\{  X\left(  t\right)  ,t\geq0\right\}  $ with (\ref{Equ7.2}) to
(\ref{Equ7.3}),\ or the uniqueness of solution to the infinite-size system of
integral-differential equations (\ref{Equ3}) to (\ref{Equ6}). To that end, a
sufficient condition is Lipschitz, that is, for some constant $M>0,$%
\[
|F\left(  y\right)  -F\left(  z\right)  |\leq M|y-z|.
\]
In general, the Lipschitz condition is standard and sufficient for the
uniqueness of solution to the finite-size system of differential equations;
while for the countable infinite-size case, readers may refer to Theorem 3.2
in Deimling \cite{Dei:1977} and Subsection 3.4.1 in Mitzenmacher
\cite{Mit:1996b} for some generalization.

To check the Lipschitz condition, as $n\rightarrow\infty$ we have
\[
\beta_{a}\left(  S_{0}\right)  =-\lambda,
\]%
\[
\beta_{b}\left(  S_{0}\right)  =\int_{0}^{+\infty}\mu\left(  x\right)
S_{1}\left(  x\right)  \text{d}x;
\]%
\[
\beta_{a}\left(  S_{1}\right)  =\lambda-\lambda\int_{0}^{+\infty}\left[
S_{1}\left(  x\right)  \right]  ^{d}\text{d}x,
\]%
\[
\beta_{b}\left(  S_{1}\right)  =-\int_{0}^{+\infty}\mu\left(  x\right)
S_{1}\left(  x\right)  \text{d}x+\int_{0}^{+\infty}\mu\left(  x\right)
S_{2}\left(  x\right)  \text{d}x;
\]
and for $k\geq2,$%
\[
\beta_{a}\left(  S_{k}\right)  =\lambda\int_{0}^{+\infty}\left[
S_{k-1}\left(  x\right)  \right]  ^{d}\text{d}x-\lambda\int_{0}^{+\infty
}\left[  S_{k}\left(  x\right)  \right]  ^{d}\text{d}x,
\]%
\[
\beta_{b}\left(  S_{k}\right)  =-\int_{0}^{+\infty}\mu\left(  x\right)
S_{k}\left(  x\right)  \text{d}x+\int_{0}^{+\infty}\mu\left(  x\right)
S_{k+1}\left(  x\right)  \text{d}x.
\]

Let%
\[
\zeta_{k}=\frac{\int_{0}^{+\infty}\left[  S_{k}\left(  x\right)  \right]
^{d}\text{d}x}{\int_{0}^{+\infty}S_{k}\left(  x\right)  \text{d}x}%
\]
and%
\[
\eta_{k}=\frac{\int_{0}^{+\infty}\mu\left(  x\right)  S_{k}\left(  x\right)
\text{d}x}{\int_{0}^{+\infty}S_{k}\left(  x\right)  \text{d}x}.
\]
Then $\zeta_{k},\eta_{k}>0$ for $k\geq1$.

The following theorem shows that the supermarket model with general service
times satisfies the Lipschitz condition for the infinite-size system of
integral-differential equations (\ref{Equ3}) to (\ref{Equ6}).

\begin{The}
\label{The:Lip}The supermarket model with general service times satisfies the
Lipschitz condition.
\end{The}

\textbf{Proof} \ Let%
\[
\Omega=\left\{  S_{k}:k\geq0\right\}  .
\]
For two arbitrary entries $y,z\in\Omega$, we have%
\[
|F\left(  y\right)  -F\left(  z\right)  |\leq a|\beta_{a}\left(  y\right)
-\beta_{a}\left(  z\right)  |+b|\beta_{b}\left(  y\right)  -\beta_{b}\left(
z\right)  |.
\]

Now, we analyze the following four cases for the function $\beta_{a}\left(
y\right)  $, while the function $\beta_{b}\left(  y\right)  $ can be analyzed similarly.

Case one: $y=S_{0},z=S_{1}$. In this case, we have%
\begin{align*}
|\beta_{a}\left(  y\right)  -\beta_{a}\left(  z\right)  |  &  =|-\lambda
-\lambda+\lambda\int_{0}^{+\infty}\left[  S_{1}\left(  x\right)  \right]
^{d}\text{d}x|\\
&  =\lambda|2-\zeta_{1}\int_{0}^{+\infty}S_{1}\left(  x\right)  \text{d}x|\\
&  =\lambda\left[  2-\zeta_{1}\int_{0}^{+\infty}S_{1}\left(  x\right)
\text{d}x\right]  ,
\end{align*}
since $0<\zeta_{1},\int_{0}^{+\infty}S_{1}\left(  x\right)  $d$x<1$. Taking%
\[
M_{a}\left(  0,1\right)  \geq\frac{2-\zeta_{1}\int_{0}^{+\infty}S_{1}\left(
x\right)  \text{d}x}{2-\int_{0}^{+\infty}S_{1}\left(  x\right)  \text{d}x},
\]
it is clear that%
\begin{align*}
|\beta_{a}\left(  y\right)  -\beta_{a}\left(  z\right)  |  &  \leq
M_{a}\left(  0,1\right)  \lambda\left[  2-\int_{0}^{+\infty}S_{1}\left(
x\right)  \text{d}x\right] \\
&  =M_{a}\left(  0,1\right)  |y-z|.
\end{align*}

Case two: $y=S_{0},z=S_{k}$ for $k\geq2$. In this case, we have%
\begin{align*}
|\beta_{a}\left(  y\right)  -\beta_{a}\left(  z\right)  |  &  =|-\lambda
-\lambda\int_{0}^{+\infty}\left[  S_{k-1}\left(  x\right)  \right]
^{d}\text{d}x+\lambda\int_{0}^{+\infty}\left[  S_{k}\left(  x\right)  \right]
^{d}\text{d}x|\\
&  =\lambda|-1-\zeta_{k-1}\int_{0}^{+\infty}S_{k-1}\left(  x\right)
\text{d}x+\zeta_{k}\int_{0}^{+\infty}S_{k}\left(  x\right)  \text{d}x|\\
&  =\lambda\left[  1+\zeta_{k-1}\int_{0}^{+\infty}S_{k-1}\left(  x\right)
\text{d}x-\zeta_{k}\int_{0}^{+\infty}S_{k}\left(  x\right)  \text{d}x\right]
\end{align*}
due to that $0<\zeta_{k},\int_{0}^{+\infty}S_{k}\left(  x\right)  $d$x<1$. Let%
\[
M_{a}\left(  0,k\right)  \geq\frac{1+\zeta_{k-1}\int_{0}^{+\infty}%
S_{k-1}\left(  x\right)  \text{d}x-\zeta_{k}\int_{0}^{+\infty}S_{k}\left(
x\right)  \text{d}x}{1+\int_{0}^{+\infty}S_{k-1}\left(  x\right)
\text{d}x-\int_{0}^{+\infty}S_{k}\left(  x\right)  \text{d}x.}%
\]
Then%
\begin{align*}
|\beta_{a}\left(  y\right)  -\beta_{a}\left(  z\right)  |  &  \leq
M_{a}\left(  0,k\right)  \lambda\left[  1+\int_{0}^{+\infty}S_{k-1}\left(
x\right)  \text{d}x-\int_{0}^{+\infty}S_{k}\left(  x\right)  \text{d}x\right]
\\
&  =M_{a}\left(  0,k\right)  |y-z|.
\end{align*}

Case three: $y=S_{1},z=S_{k}$ for $k\geq2$. In this case, we have%
\begin{align*}
|\beta_{a}\left(  y\right)  -\beta_{a}\left(  z\right)  |  &  =|\lambda
-\lambda\int_{0}^{+\infty}\left[  S_{1}\left(  x\right)  \right]  ^{d}%
\text{d}x-\lambda\int_{0}^{+\infty}\left[  S_{k-1}\left(  x\right)  \right]
^{d}\text{d}x+\lambda\int_{0}^{+\infty}\left[  S_{k}\left(  x\right)  \right]
^{d}\text{d}x|\\
&  =\lambda|1-\zeta_{1}\int_{0}^{+\infty}S_{1}\left(  x\right)  \text{d}%
x-\zeta_{k-1}\int_{0}^{+\infty}S_{k-1}\left(  x\right)  \text{d}x+\zeta
_{k}\int_{0}^{+\infty}S_{k}\left(  x\right)  \text{d}x|
\end{align*}
Let%
\[
M_{a}\left(  1,k\right)  \geq\frac{|1-\zeta_{1}\int_{0}^{+\infty}S_{1}\left(
x\right)  \text{d}x-\zeta_{k-1}\int_{0}^{+\infty}S_{k-1}\left(  x\right)
\text{d}x+\zeta_{k}\int_{0}^{+\infty}S_{k}\left(  x\right)  \text{d}%
x|}{|1-\int_{0}^{+\infty}S_{1}\left(  x\right)  \text{d}x-\int_{0}^{+\infty
}S_{k-1}\left(  x\right)  \text{d}x+\int_{0}^{+\infty}S_{k}\left(  x\right)
\text{d}x|}.
\]
Then%
\begin{align*}
|\beta_{a}\left(  y\right)  -\beta_{a}\left(  z\right)  |  &  \leq
M_{a}\left(  1,k\right)  \lambda|1-\int_{0}^{+\infty}S_{1}\left(  x\right)
\text{d}x-\int_{0}^{+\infty}S_{k-1}\left(  x\right)  \text{d}x+\int
_{0}^{+\infty}S_{k}\left(  x\right)  \text{d}x|\\
&  =M_{a}\left(  1,k\right)  |y-z|.
\end{align*}

Case four: $y=S_{l},z=S_{k}$ for $k>l\geq2$. In this case, we have%
\begin{align*}
|\beta_{a}\left(  y\right)  -\beta_{a}\left(  z\right)  | =  &  |\lambda
\int_{0}^{+\infty}\left[  S_{l-1}\left(  x\right)  \right]  ^{d}%
\text{d}x-\lambda\int_{0}^{+\infty}\left[  S_{l}\left(  x\right)  \right]
^{d}\text{d}x\\
&  -\lambda\int_{0}^{+\infty}\left[  S_{k-1}\left(  x\right)  \right]
^{d}\text{d}x+\lambda\int_{0}^{+\infty}\left[  S_{k}\left(  x\right)  \right]
^{d}\text{d}x|\\
=  &  \lambda|\zeta_{l-1}\int_{0}^{+\infty}S_{l-1}\left(  x\right)
\text{d}x+\zeta_{l}\int_{0}^{+\infty}S_{l}\left(  x\right)  \text{d}x\\
&  -\zeta_{k-1}\int_{0}^{+\infty}S_{k-1}\left(  x\right)  \text{d}x+\zeta
_{k}\int_{0}^{+\infty}S_{k}\left(  x\right)  \text{d}x|.
\end{align*}
Let%
\[
M_{a}\left(  l,k\right)  \geq\frac{|\zeta_{l-1}\int_{0}^{+\infty}%
S_{l-1}\left(  x\right)  \text{d}x+\zeta_{l}\int_{0}^{+\infty}S_{l}\left(
x\right)  \text{d}x-\zeta_{k-1}\int_{0}^{+\infty}S_{k-1}\left(  x\right)
\text{d}x+\zeta_{k}\int_{0}^{+\infty}S_{k}\left(  x\right)  \text{d}x|}%
{|\int_{0}^{+\infty}S_{l-1}\left(  x\right)  \text{d}x+\int_{0}^{+\infty}%
S_{l}\left(  x\right)  \text{d}x-\int_{0}^{+\infty}S_{k-1}\left(  x\right)
\text{d}x+\int_{0}^{+\infty}S_{k}\left(  x\right)  \text{d}x|}.
\]
Then%
\[
|\beta_{a}\left(  y\right)  -\beta_{a}\left(  z\right)  |\leq M_{a}\left(
l,k\right)  |y-z|.
\]

Based on the above four cases, taking%
\[
M_{a}=\max\left\{  M_{a}\left(  l,k\right)  :k>l\geq0\right\}
\]
we obtain that for two arbitrary entries $y,z\in\Omega,$%
\[
|\beta_{a}\left(  y\right)  -\beta_{a}\left(  z\right)  |\leq M_{a}|y-z|.
\]

Similarly, we can choose a positive number $M_{b}$ such that for two arbitrary
entries $y,z\in\Omega,$%
\[
|\beta_{b}\left(  y\right)  -\beta_{b}\left(  z\right)  |\leq M_{b}|y-z|.
\]

Let $M=\max\left\{  aM_{a},bM_{b}\right\}  $. Then for two arbitrary entries
$y,z\in\Omega,$%
\[
|F\left(  y\right)  -F\left(  z\right)  |\leq M|y-z|.
\]
This completes the proof. \nopagebreak   \hspace*{\fill}
\nopagebreak                   \textbf{{\rule{0.2cm}{0.2cm}}}

Based on Theorem \ref{The:Lip}, the following theorem easily follows from
Theorem 3.13 in Mitzenmacher \cite{Mit:1996b}.

\begin{The}
\label{The:Kurtz}In the supermarket model with general service times,
$\left\{  X_{n}\left(  t\right)  \right\}  $ and $\left\{  X\left(  t\right)
\right\}  $ are respectively given by (\ref{Equ7.5}) and (\ref{Equ7.6}), we
have%
\[
\lim_{n\rightarrow\infty}\sup_{u\leq t}|X_{n}\left(  u\right)  -X\left(
u\right)  |=0,\text{ \ }a.s.
\]
\end{The}

\textbf{Proof} \ It is seen from that in the supermarket model with general
service times, the function $F\left(  y\right)  $ for $y\in\Omega$ satisfies
the Lipschitz condition. At the same time, it is easy to take a subset
$\Omega^{\ast}\subset\Omega$ such that%
\[
\left\{  X\left(  u\right)  :u\leq t\right\}  \subset\Omega^{\ast}%
\]
and%
\[
a\sup_{y\in\Omega^{\ast}}\beta_{a}\left(  y\right)  +a\sup_{y\in\Omega^{\ast}%
}\beta_{a}\left(  y\right)  <+\infty.
\]
Thus, this proof can easily be completed by means of Theorem 3.13 in
Mitzenmacher \cite{Mit:1996b}. This completes the proof. \nopagebreak
\hspace*{\fill} \nopagebreak        \textbf{{\rule{0.2cm}{0.2cm}}}

Using Theorem 3.11 in Mitzenmacher \cite{Mit:1996b} and Theorem
\ref{The:Kurtz}, we can obtain the following theorem for the expected sojourn
time that a customer spends in an initially empty supermarket model with
general service times over the time interval $\left[  0,T\right]  $.

\begin{The}
In the supermarket model with general service times, the expected sojourn time
that a customer spends in an initially empty system over the time interval
$\left[  0,T\right]  $ is bounded above by%
\[
\theta\rho^{d}\left\{  E\left[  X_{R}\right]  -E\left[  X\right]  \right\}
+E\left[  X\right]  \left[  \sum_{k=1}^{\infty}\theta^{\frac{d^{k}-1}{d-1}%
}\rho^{\frac{d^{k}-d}{d-1}}\right]  +o\left(  1\right)  ,
\]
where $o\left(  1\right)  $ is understood as $n\rightarrow\infty$.
\end{The}

\section{Concluding remarks}

In this paper, we provide a novel and simple approach to study the randomized
load balancing model with general service times, which is described as an
infinite-size system of integral-differential equations. This approach is
based on the supplementary variable method, which is always applied in dealing
with stochastic models of M/G/1 type, e.g., see Li and Zhao \cite{Li:2004,
Li:2006} and Li \cite{Li:2010}. We organize an infinite-size system of
integral-differential equations by means of the density dependent jump Markov
process, and obtain a close-form solution: doubly exponential structure, for
the fixed point satisfying the system of nonlinear equations, which is always
a key in the study of supermarket models. Since the fixed point is
decomposited into two groups of information under a product form, we indicate
three important observations:

\begin{enumerate}
\item[1.] the fixed point for the supermarket model is different from the tail
of stationary queue length distribution for the ordinary M/G/1 queue;

\item[2.] the doubly exponential solution to the fixed point can exist
extensively for $0<\mu<+\infty$ even if the service time distribution is heavy-tailed; and

\item[3.]  the doubly exponential solution to the fixed point is not unique for a
more general supermarket model.
\end{enumerate}

\noindent Furthermore, we analyze the exponential convergence of the current
location of the supermarket model to its fixed point, and study the Lipschitz
condition in the Kurtz Theorem under general service times. Finally, we
present numerical examples to illustrate the effectiveness of our approach in
analyzing the randomized load balancing schemes with the non-exponential
service requirements. Based on this analysis, one can gain a new and important
understanding how workload probing can help in load balancing jobs with
general service times such as heavy-tailed service.

The approach of this paper is useful in analyzing the randomized load
balancing schemes in resource allocation in computer networks. We expect that
this approach will be applicable to the study other randomized load balancing
schemes with general service times, for example, generalizing the arrival
process to non-Poisson: the renewal arrival process or the Markovian arrival process.

\section*{Acknowledgements}

The work of Q.L. Li was supported by the National Science Foundation of China
under grant No. 10871114 and the National Grand Fundamental Research 973
Program of China under grant No. 2006CB805901.

\end{document}